\documentclass[aps,prd,showpacs,
twocolumn
,superscriptaddress]{revtex4}
\usepackage{graphicx}
\usepackage{dcolumn}
\usepackage{bm}
\usepackage{color}
\usepackage[normalem]{ulem} 
\usepackage[dvipsnames]{xcolor} 
\usepackage{hyperref}
\hypersetup{
  colorlinks=true,        
  linkcolor=blue,         
  citecolor=cyan,         
}
\usepackage{mathrsfs}
\usepackage[utf8]{inputenc}
\usepackage{mathtools}
\usepackage{doi}
\usepackage{amsmath}
\usepackage{amssymb}

\begin{document}

\title{Dynamics of test particles around regular black holes in modified gravity}
\author{Javlon Rayimbaev}
\email{javlon@astrin.uz}
\affiliation{Ulugh Beg Astronomical Institute, 33 Astronomy st., Tashkent 100052, Uzbekistan}
\affiliation{National University of Uzbekistan, Tashkent 100174, Uzbekistan}
\affiliation{Institute of Nuclear Physics, Ulugbek 1, Tashkent 100214, Uzbekistan}
\author{Pulat~Tadjimuratov}
\email{tadjimura@astrin.uz}
\affiliation{Ulugh Beg Astronomical Institute, 33 Astronomy st., Tashkent 100052, Uzbekistan}
\affiliation{Webster University in Tashkent, 13 Navoiy st., Tashkent 100011, Uzbekistan}
\author{Ahmadjon~Abdujabbarov}
\email{ahmadjon@astrin.uz}
\affiliation{Shanghai Astronomical Observatory, 80 Nandan Road, Shanghai 200030, P. R. China}
\affiliation{Ulugh Beg Astronomical Institute, 33 Astronomy st., Tashkent 100052, Uzbekistan}
\affiliation{National University of Uzbekistan, Tashkent 100174, Uzbekistan}
\affiliation{Institute of Nuclear Physics, Ulugbek 1, Tashkent 100214, Uzbekistan}
\affiliation{Tashkent Institute of Irrigation and Agricultural Mechanization Engineers, 39 Kori Niyoziy st., Tashkent 100000, Uzbekistan}
\author{Bobomurat~Ahmedov}
\email{ahmedov@astrin.uz}
\affiliation{Ulugh Beg Astronomical Institute, 33 Astronomy st., Tashkent 100052, Uzbekistan} \affiliation{National University of Uzbekistan, Tashkent 100174, Uzbekistan} 
\affiliation{Tashkent Institute of Irrigation and Agricultural Mechanization Engineers, 39 Kori Niyoziy st., Tashkent 100000, Uzbekistan}

\author{Malika~Khudoyberdieva}
\email{xudoyberdiyeva94@inbox.ru}
\affiliation{National University of Uzbekistan, Tashkent 100174, Uzbekistan}

\date{\today}

\begin{abstract}

In the work, we have presented detailed analyses of the event horizon and curvature properties of spacetime around a regular black hole in modified gravity so-called regular MOG black hole. The motion of neutral, electrically charged and magnetized particles with magnetic dipole moment, in the close environment of the regular MOG black hole immersed in an external asymptotically uniform magnetic field has been also explored. Through the study on the motion of the neutral test particle, we have obtained that the positive (negative) values of the MOG parameter mimic the spin of a rotating Kerr black hole giving the same values for innermost stable pro-grade (retrograde) orbits of the particles in the range of the spin parameter $a/M \in(-0.4125, \ 0.6946)$. The efficiency of energy release from the accretion disk by the Novikov-Thorne model has been calculated and shown that the efficiency is linearly proportional to the increase of the MOG parameter $\alpha$, and exceeds up to 15 \% at the critical limiting value of the MOG parameter $\alpha_{\rm cr}$.  The study of the charged particles dynamics has shown that innermost stable circular orbits (ISCOs) of the particle increases with the increase of the MOG parameter, while the increase of cyclotron frequency being responsible to the magnetic interaction causes the decrease of the latter. Moreover, the dynamics of magnetic dipoles has shown that the increase of the MOG and the magnetic coupling parameters lead to an increase of the inner radius and the width of the accretion disk consisting of magnetized particles. Finally, we have focused on showing the range of values of the magnetic coupling parameter causing the orbits to be stable.

\end{abstract}
\pacs{04.50.-h, 04.40.Dg, 97.60.Gb}

\maketitle

\section{Introduction}

The vicinity of the astrophysical black holes being compact gravitating object may serve as laboratory for testing the gravity theories in the strong field regime. The test of gravity theories through X-ray observation of stellar mass (3-30 $M_\odot$) and supermassive ( up to $10^{10}~M_\odot$) has been carried out by numerous authors (see, e.g. \cite{Bambi17c}). Other ways of testing the gravity in strong field regime are through detection of gravitational waves by  LIGO/VIRGO~\cite{LIGO} and image of supermassive black holes at the Event Horizon Telescope (EHT)~\cite{EHT19a,EHT19b}. 

The first spherical symmetric vacuum solution within the general relativity describing the nonrotating black hole with total mass $M$ has been been obtained by Schwarzschild in 1916~\cite{Schwarzschild16a}. Later Reissner and Nordstr\"om have independently found the solution describing electrically charged  black hole~\cite{Reissner16,Nordstrom18}. The fundamental problem of these and other exact solutions of the field equations in general relativity is the presence of the singularity at the center of  black hole, which can not be resolved by the concept of the classical theory. Regular black hole solutions obtained by coupling general relativity to nonlinear electrodynamics obeying the singularity problem are described in e.g. Refs.~\cite{Bardeen:1970,Hayward04,Dymnikova11,Dymnikova15,Toshmatov17a,Toshmatov17,Bronnikov00,Toshmatov18b,Toshmatov14,Toshmatov15b,Stuchlik15,Schee15,Toshmatov2018PhRvD2,Toshmatov2018PhRvD,Toshmatov2019PhRvD..99f4043T}. Regular black holes in modified gravity have been investigated in~\cite{Moffat2017EPJP}.   

The spacetime around black holes is strongly curved leading to the formation of event horizon which is a surface bounding the region of spacetime unobservable by the external observer. Depending on the central black hole the spacetime geometry near the horizon differs from each other. The  geometrical structure of the spacetime can be determined by the motion of the particles in the vicinity of the black hole~\cite{Jawad16, Hussain15, Babar16, Banados09, Majeed17, Zakria15, Brevik19}. Existence of a magnetic field around black hole also alters the particle motion resulting in the chaotic dynamics~\cite{Chen16, Hashimoto17, Dalui19, Han08, Moura00}. 

The existence of the dark energy and dark matter in the Universe has been proven by the several observations (e.g. the galactic rotation curves and the accelerated expansion of the Universe, etc.), which can be explained with introducing the cosmological constant $\Lambda$ in the cosmological solutions of general relativity and hidden mass as cold dark matter so-called standard $\Lambda$CDM cosmological model. However, there are modifications of general relativity that try to explain such phenomena by the intrinsic effects of the extended theories of gravity. One of such modifications is so-called  Scalar-Tensor-Vector Gravity (STVG, known also as MOG) proposed by Moffat~\cite{Moffat06}. The theory has been applied to galaxy rotation curves~\cite{Moffat13, Moffat15b}, black hole shadow~\cite{Moffat15, Moffat15a}, supernovae~\cite{Wondrak18} and gravitational lensing~\cite{Moffat09}. Non rotating and rotating black hole  solutions in MOG (known as Schwarzschild-MOG and Kerr-MOG ones, respectively) have been obtained in~\cite{Moffat15}.

The no-hair theorem does not allow the black holes to have their own magnetic field, however black holes can be considered in external magnetic field generated, e.g. by either charged particles in accretion disc around the black hole or magnetized companion star  of the black hole in close binary systems. The solution of electromagnetic field equation around black hole embedded in an external asymptotically uniform magnetic field has been first obtained by Wald~\cite{Wald74} and various observational properties around the black hole in external asymptotically uniform magnetic field in strong gravity regime in the different theories of gravity have been developed by several authors~\cite{Aliev86,Aliev89,Aliev02,Frolov11,Frolov12,Benavides-Gallego18,Shaymatov18,Stuchlik14a,Abdujabbarov10,Abdujabbarov11a,Abdujabbarov11,Abdujabbarov08,Karas12a,Stuchlik16,Kovar10,Kovar14,Kolos17,Pulat2020PhRvDMOG,Rayimbaev2019IJMPCS,Rayimbaev2020MPLA,Rayimbaev2019IJMPD}.
Investigation of the dynamics of particles with magnetic dipole and electrically as well as magnetically charged particles around black hole in external magnetic field and black hole with electric and magnetic charge have been discussed in detail in the framework of various alternative theories of gravity ~\cite{deFelice,deFelice2004, Rayimbaev16, Oteev16,Toshmatov15d,Abdujabbarov14,Rahimov11a,Rahimov11,Haydarov20,Haydarov2020EPJC,Abdujabbarov2020PDU,Narzilloev2020PhRvDstringy,Rayimbaev2020PhRvD,TurimovPhysRevD2020,DeLaurentis2018PhRvD,MorozovaV2014PhRvD,Nathanail2017MNRAS,Vrba2020PhRvD,Vrba2019EPJC,Morozova2014PhRvD}. 

In this work we have studied the test neutral, electrically charged and magnetized particles motion around regular MOG black hole immersed in an asymptotically uniform external magnetic field. Sect.~\ref{chapter1} focuses on the analysis of the properties of spacetime geometry around regular MOG black hole, in Sects.~\ref{testpartmotion}, \ref{chargedpartmotion} and \ref{magnetpartmotion} we investigate the dynamics of neutral, charged and magnetized particles, respectively. Finally, in Sect.~\ref{conclusion} we summarize the obtained main results. Throughout the paper a spacelike signature $(-,+,+,+)$  is selected  for the spacetime and the geometrized system of units is used where $G = c= h = 1$ (However, for an astrophysical application the speed of light and the Newtonian gravitational constant are written explicitly in our expressions). The Latin indices run from $1$ to $3$ and the Greek ones from $0$ to $3$.

\section{The spacetime properties \label{chapter1}}

In this section we explore geometric properties of spherical symmetric spacetime around regular MOG black holes such as curvature invariants and the existence of event horizon. The spacetime around the regular MOG black hole can be described by the following line element~\cite{Moffat15,Mureika16}.
\begin{eqnarray}\label{metric}
ds^2=-f(r)dt^2+\frac{1}{f(r)}dr^2+r^2 (d\theta^2 +\sin^2\theta d\phi^2)\ ,
\end{eqnarray}
with  
\begin{eqnarray} \label{metfunct}\nonumber
f(r)&=&1-\frac{2M}{r}\frac{\alpha+1}{\left[ 1+\alpha  (\alpha +1) \frac{M^2}{r^2} \right]^{3/2}}\\
&&+\frac{\alpha  (\alpha +1) \frac{M^2}{r^2}}{\left[ 1+\alpha  (\alpha +1) \frac{M^2}{r^2} \right]^{2}}\ , 
\end{eqnarray}
where $M$ is the total mass of the black hole, $\alpha$ is so-called MOG parameter being responsible for the modified gravity. The metric (\ref{metric}) takes the form of the Schwarzschild solution when $\alpha=0$ and at $\alpha=-1$ the metric reflects flat spacetime effects. There are several constrains for the parameter $\alpha$ obtained by Moffat et.al using the different astronomical observational data. For example, in Ref.\cite{Moffat2017EPJP} authors have shown that MOG plays the role of dark matter providing the same gravitational redshift with the MOG parameter $\alpha=8.89$ based on the data from X-ray band observations of the galactic cluster Abel 1689. Analyzing the data from the gravitational wave events GW150914 and GW151226 authors of Ref.\citep{Moffat2016PhLB} have obtained the values of the MOG parameter in the range $\alpha \to (2 \div 8.3)$. It has been shown in Ref.~\cite{Moffat2019MNRAS} that the effect of MOG near the  Sgr A* is weak  and the upper limit of the  parameter takes the value $\alpha=0.055$. Other constraint for MOG parameter as $\alpha=1.13^{+0.30}_{-0.24}$ has been obtained using the observational data on image of M87 event horizon ~\cite{Moffat2020PhRvD}. The analysis of observational data from rotation curves of the nearby galaxies has allowed to get constraint on upper value of MOG parameter as $\alpha=8.89 \pm 0.34$~\cite{Moffat2013MNRAS}.

\subsection{Scalar invariants}\label{RRKinvariants}
 
Here we plan to study the curvature invariants such as the Ricci scalar, square of the Ricci tensor and the Kretschmann scalar. It may help to deeply understand the main properties of the spacetime~(\ref{metric}). 

{\bf The Ricci scalar. }
The expression for the Ricci scalar can be easily derived in the following form 
\begin{eqnarray}\label{R}
 R&=&g^{\mu \nu}R_{\mu \nu}=\nonumber \\ &=& \frac{6 \alpha  (\alpha +1)^2 M^3 }{\mathcal{A}^9 r}\Bigg\{4 \alpha ^2 (\alpha +1)^2 \frac{M^4}{r^4}+3 \alpha (\alpha +1) \frac{M^2}{r^2}\nonumber\\ 
 &&+2 \alpha  \mathcal{A} \frac{M}{r} \left[1-\alpha  (\alpha +1) \frac{M^2}{r^2}\right]-1\Bigg\}\ ,
\end{eqnarray}
where $\mathcal{A}=\sqrt{1+\alpha  (\alpha +1)M^2/r^2}$. Analyzing the Eq.~(\ref{R}) near the central point one can easily see that there is no singularity and the solution is regular at the center: 
\begin{eqnarray}\label{R1}
 \lim_{r \to 0}R= \frac{12 M^{-2} }{\alpha ^{3/2} (\alpha +1)^{3/2}}\left[2 (\alpha +1)-\sqrt{\alpha  (\alpha +1)}\right]\ ,
\end{eqnarray}

{\bf The square of Ricci tensor. }
Now we explore the square of Ricci tensor of spacetime (\ref{metric}) which can be found in the following form 
 \begin{eqnarray}\label{RR}
 \nonumber 
 && R_{\mu \nu} R^{\mu \nu} = \frac{2 \alpha ^2 (\alpha +1)^2 M^4}{\mathcal{A}^{16} r^8} \Bigg\{2-\frac{72 \alpha ^3 (\alpha +1)^4 \mathcal{A} M^7}{r^7}\\\nonumber && +\frac{90 \alpha ^2 (\alpha +1)^3 \mathcal{A} M^5}{r^5}-\frac{168 \alpha  (\alpha +1)^2 \mathcal{A} M^3}{r^3}\\\nonumber &&  +\frac{30 (\alpha +1) \mathcal{A} M}{r}+\frac{18 \alpha ^3 (\alpha +1)^4 (5 \alpha +4) M^8}{r^8}\\\nonumber && +\frac{36 \alpha ^2 (\alpha +1)^3 M^6}{r^6}
 +\frac{\alpha  (\alpha +1)^2 (149 \alpha +81) M^4}{r^4}\\ && +\frac{(\alpha +1) (97 \alpha +117) M^2}{r^2}\Bigg\}
\end{eqnarray}
One may easily see that the square of the Ricci tensor is also finite, as well as Ricci scalar. At the center it tkes the following form: 
\begin{eqnarray}\label{RR1}
 \lim_{r \to 0}R_{\mu \nu} R^{\mu \nu}=\frac{36 M^{-4}}{\alpha ^3 (\alpha +1)^2}\left[5 \alpha -4 \sqrt{\alpha(\alpha +1)}+4\right] \ .
 \end{eqnarray}

{\bf The Kretschmann scalar. }
Now we explore  the Kretschmann scalar, which gives more information about the curvature of the spacetime (\ref{metric}), since the Kretschmann scalar does not vanish even for Ricci flat spacetime and is helpful to study the properties of a given Ricci flat spacetime. The exact analytic expression for the Kretschmann scalar with the spacetime of the metric (\ref{metric}) is
\begin{eqnarray}\label{K}
  K&=&R_{\mu \nu \sigma \rho}R^{\mu \nu \sigma \rho}=\frac{4 (\alpha +1)^2 M^2}{\mathcal{A}^{16} r^8}\\\nonumber &\times & \Bigg\{12-\frac{24 \alpha ^5 (\alpha +1)^4 \mathcal{A} M^9}{r^9}+\frac{30 \alpha ^4 (\alpha +1)^3 \mathcal{A} M^7}{r^7}\\\nonumber &-& \frac{192 \alpha ^3 (\alpha +1)^2 \mathcal{A} M^5}{r^5}
+\frac{90 \alpha^2 (\alpha +1) \mathcal{A} M^3}{r^3}\\\nonumber &-& \frac{24 \alpha  \mathcal{A} M}{r}+\frac{6 \alpha ^5 (\alpha +1)^4 (5 \alpha +4) M^{10}}{r^{10}}\\\nonumber &+& \frac{12 \alpha ^4 (\alpha +1)^3 M^8}{r^8}+\frac{\alpha ^3 (\alpha +1)^2 (197 \alpha +129) M^6}{r^6}\\\nonumber &+& \frac{\alpha ^2 (\alpha +1) (61 \alpha +105) M^4}{r^4}-\frac{2 \alpha  (5 \alpha +12) M^2}{r^2}\Bigg\}
  \end{eqnarray}
which has the following value at the center:
 \begin{eqnarray}\label{K1}
  \lim_{r \to 0}K=\frac{24 M^{-4}}{\alpha ^3 (\alpha +1)^2 } \left[5 \alpha -4 \sqrt{\alpha  (\alpha +1)}+4\right] \ .
  \end{eqnarray}
\begin{figure}[ht!]
   \centering
  \includegraphics[width=0.9\linewidth]{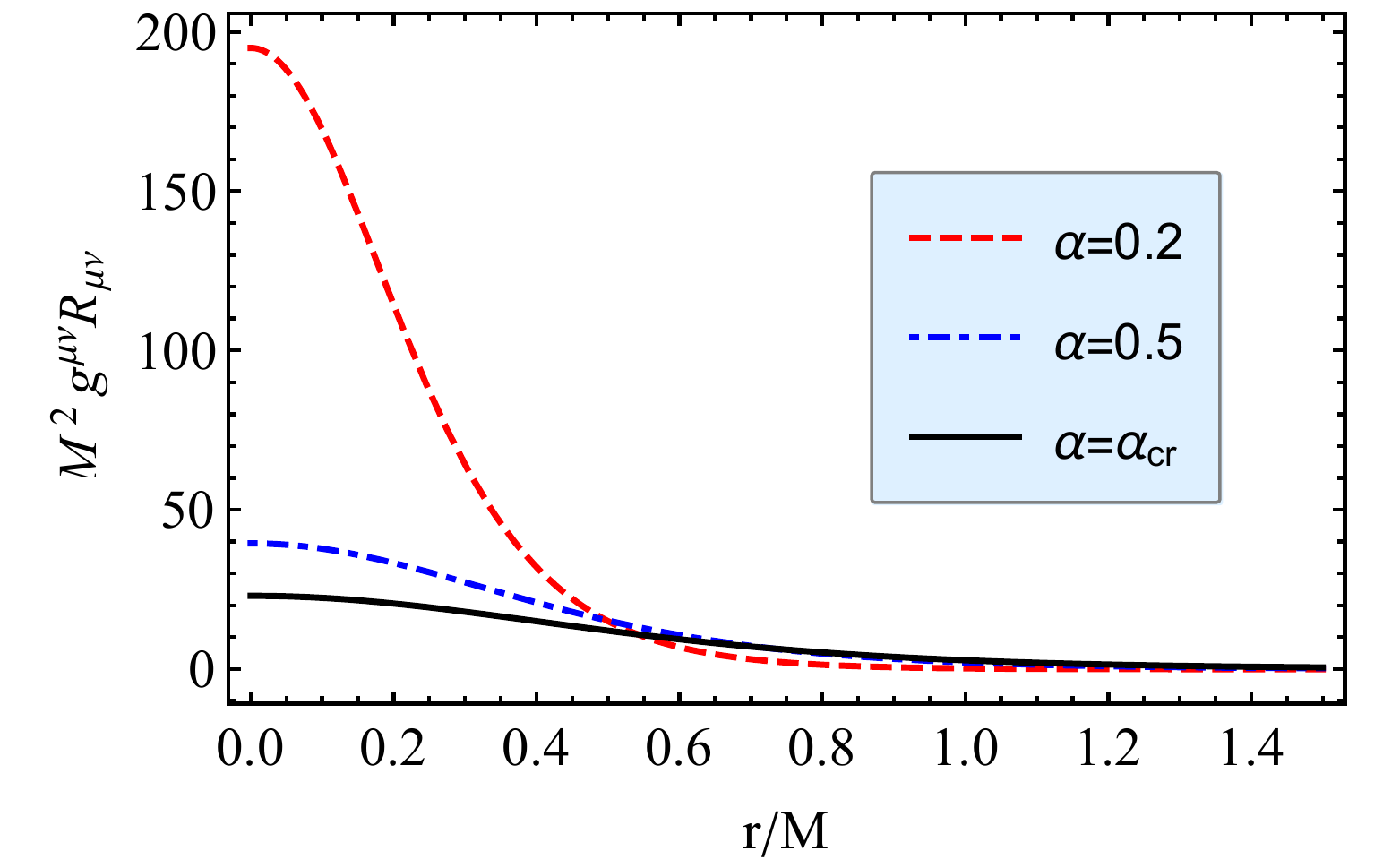}
   \includegraphics[width=0.9\linewidth]{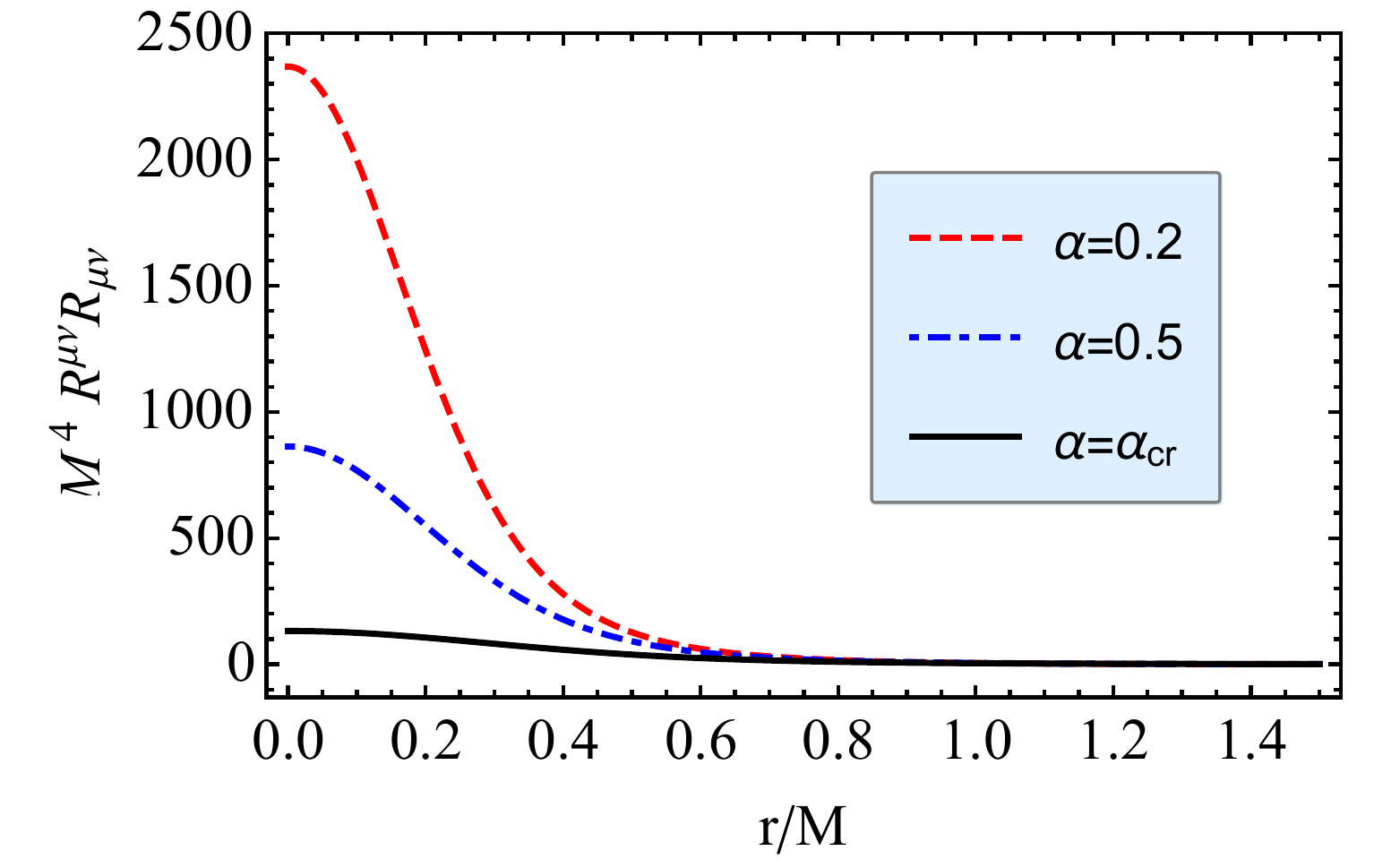}
    \includegraphics[width=0.9\linewidth]{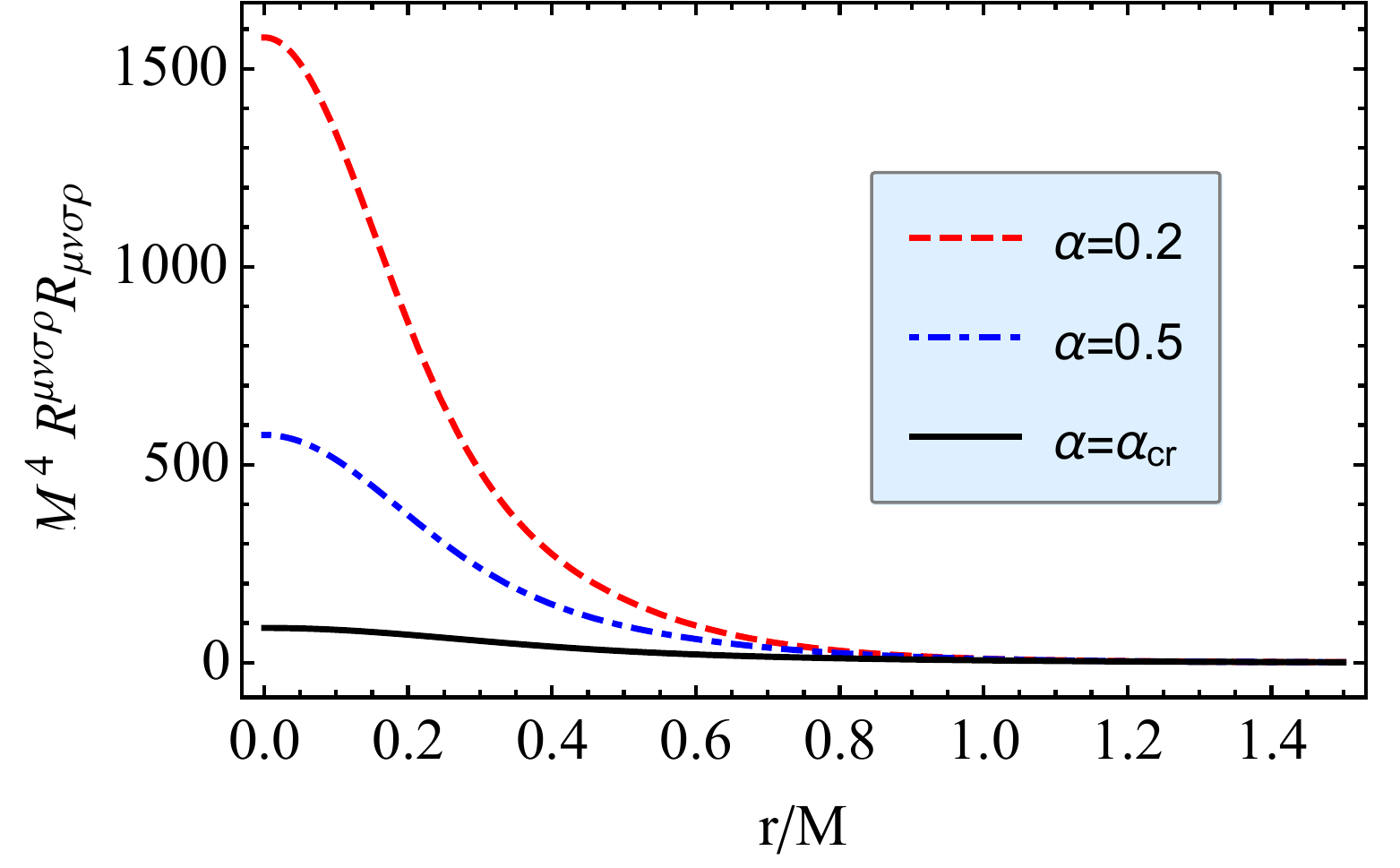}
	\caption{Radial dependence of scalar invariants of the spacetime (\ref{metric}) for the different values of the MOG parameters $\alpha$. Top panel corresponds to Ricci scalar, middle panel to square of Ricci tensor and bottom one to Kretchmann scalar. \label{ricci}}
\end{figure} 
Fig.~\ref{ricci} demonstrates the radial dependence of scalar invariants of the spacitime (\ref{metric}) for different values of the MOG parameter $\alpha$. One can see from the figure that the increase of the parameter MOG scalar invariants decrease at the center of the black hole.     

\subsection{Event horizon} 

Next we explore the event horizon properties of the Regular MOG balck hole governed by the lapse function (\ref{metfunct}) with nonvanishing MOG parameter $\alpha$.
Generally, the radius of the event horizon of a BH is defined in a standard way imposing $g_{rr} \to \infty, \ g^{rr}=0$ or equivalently, by solving the equation $f(r)=0$  with respect to $r$. We get quite complicated form of expression for radius of outer and inner horizons denoted with  $(+)$ and $(-)$ signs, respectively, in the following form
\begin{eqnarray}\label{horizoneq}
 \nonumber
 \left(\frac{r_h^{\pm}}{M}\right)^2&=&2+\alpha-\alpha ^2+ {\cal P}_2 \\ &\pm & \sqrt{{\cal P}_5+\frac{{\cal P}_6}{4 {\cal P}_2}-\frac{{\cal P}_3}{3 {\cal P}_1}-\frac{{\cal P}_1}{3 \sqrt[3]{2}}}
 \end{eqnarray}
 
 \begin{eqnarray}
 \nonumber
 {\cal P}_1^3&=& 2 \alpha ^3 (\alpha +1)^6 \Big[\alpha  (\alpha  (253 \alpha +312)+120)-64 \\\nonumber & + & 24 \sqrt{3\alpha} (\alpha +1)\sqrt{\alpha  (2 \alpha  (14 \alpha +9)-1)-16} \Big] \\\nonumber {\cal P}_2^2&=& \frac{{\cal P}_3}{3  {\cal P}_1}+\frac{ {\cal P}_1}{3 \sqrt[3]{2}}+ {\cal P}_4 \\\nonumber {\cal P}_3 &=& \sqrt[3]{2} \alpha ^2 (\alpha +1)^4 \left(25 \alpha ^2+16 \alpha +16\right) \\\nonumber
  {\cal P}_4&=&\frac{{\cal P}_5}{2}= (\alpha +1)^2 (4-\frac{\alpha}{3}(11 \alpha +4)) \\\nonumber
  {\cal P}_6&=& 32 (\alpha +1)^4 (2-3 \alpha)
 \end{eqnarray}
 
 \begin{figure}[h!]
   \centering
  \includegraphics[width=0.93\linewidth]{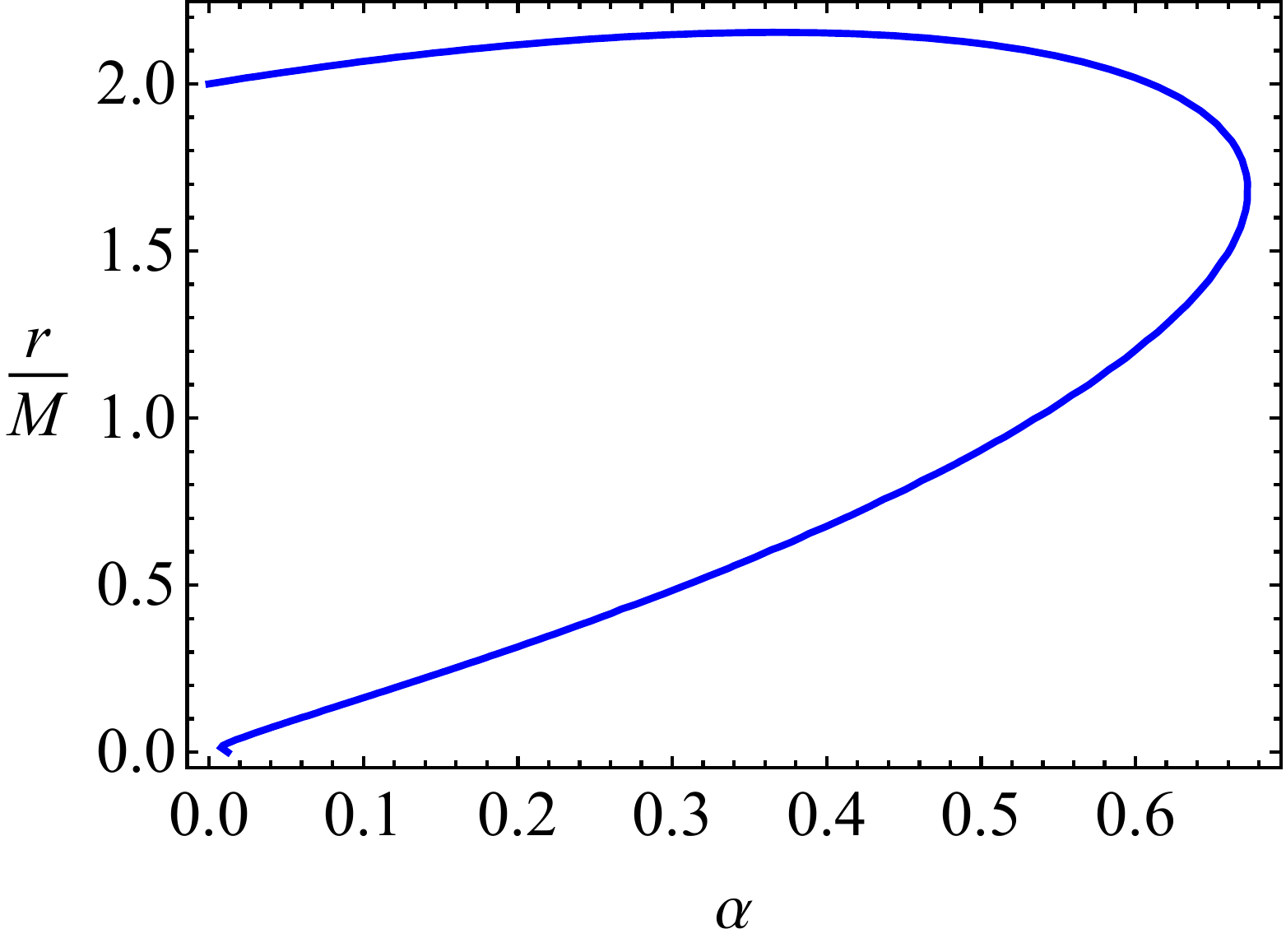}
	\caption{Dependence of event horizon radii from MOG parameter. \label{horizon}}
\end{figure} 

The dependence of the radii of event horizons for the different values of $\alpha$ is shown in Fig.~\ref{horizon}. 
 
The extreme values of the MOG parameter coincide the two outer and inner horizons. The minimum value of the outer horizon can be easily found by the following system of equations and solving them with respect to $r$ and $\alpha$
\begin{equation}
f(r)=0=f'(r)\ ,
\end{equation}

One may solve the equations of system with respect to the MOG parameter $\alpha$ and  the radial coordinate $r$  which implies the critical  value of the MOG parameter and minimum value of outer horizon. Then we have $(r_h)_{\rm min}/M=1.68119$ and $\alpha_{\rm cr}=0.67276$. 
    
Now  we explore the radial profiles of the lapse function (\ref{metfunct}) for the different values of the MOG parameter. 
    
    \begin{figure}[h!]
   \centering
  \includegraphics[width=0.98\linewidth]{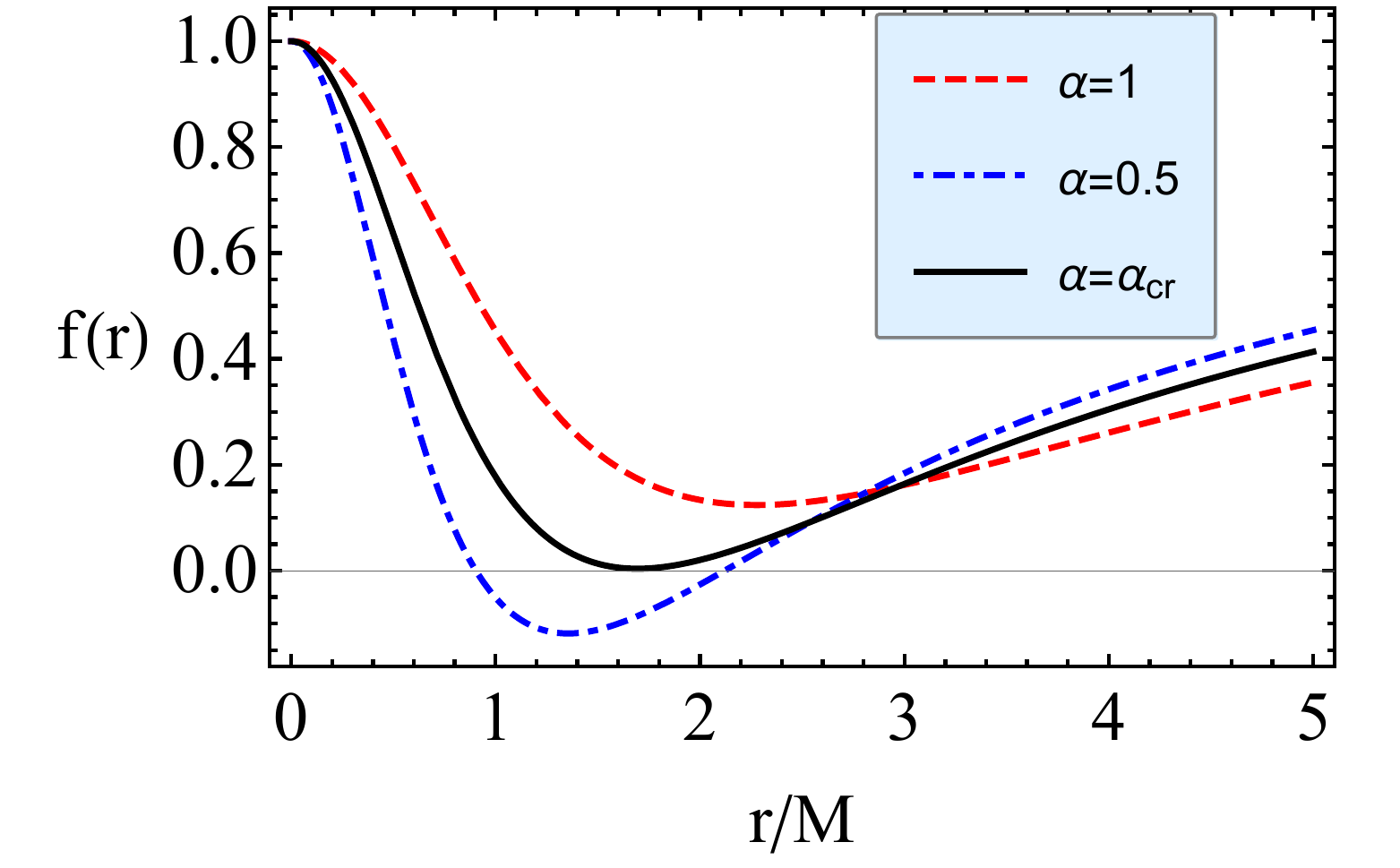}
	\caption{Dependence of lapse function on the radial coordinate $r/M$ for the different values of the parameters $\gamma$ and $\lambda$. \label{lapse}}
\end{figure} 
Figure~\ref{lapse} shows the dependence of outer and inner horizons of regular MOG black holes from the MOG parameters. One can see that the values of outer and inner horizons matches when MOG parameter takes the critical value, Consequently, we may conclude the follows
\begin{itemize}
    \item in the case when $\alpha<\alpha_{\rm cr}$ two event horizons do exist: inner and outer ones;
    \item in the case  when $\alpha>\alpha_{\rm cr}$  there are no event horizons;
    \item if $\alpha=\alpha_{\rm cr}$  we have extreme Regular MOG black hole and the two horizons coincide (see Fig.~\ref{lapse}).
\end{itemize}.

\section{Test particle motion \label{testpartmotion}}
 
In this section we will perform a detailed analysis of dynamics of test particles around regular MOG black hole described by the metric~(\ref{metric}).

\subsection{Equation of motion} 

Lagrangian density for a neutral particle with mass $m$ can be expressed in the following standard form 
\begin{eqnarray}
\mathscr{L}_{\rm p}=\frac{1}{2} g_{\mu\nu} \dot{x}^{\mu} \dot{x}^{\nu} ,
\end{eqnarray}
where $m$ is the mass of the particle. Then integrals of motion corresponding to Killing symmetries of the spacetime are read 
\begin{eqnarray}
\label{consts1}
&&  g_{tt}\dot{t}=-{\cal E}\ ,
\\\label{consts2}
&&  g_{\phi \phi}\dot{\phi} = l\ ,
\end{eqnarray}
where ${\cal E}=E/m$ and  $l=L/m$  are specific energy and specific angular momentum of the particle, respectively. One can govern the equations of motion for test particles through the normalization condition
\begin{equation}\label{norm4vel}
g_{\mu \nu}u^{\mu}u^{\nu}=\epsilon \ ,
\end{equation}
where $\epsilon$ takes the values $0$ and $-1$ for massless and massive particles, respectively.

For the massive neutral particles motion is governed by timelike geodesics of the spacetime (\ref{metric}) and the equations of motion can be derived using Eq.(\ref{norm4vel}). Taking into account the  equations (\ref{consts1}) and (\ref{consts2}), ona may obtain the equations of motion in the separated and integrable form
\begin{eqnarray}\label{eqmotionneutral}
\dot{r}^2&=&{\cal E}^2+g_{tt}\left(1+\frac{\cal K}{r^2}\right)\ ,
 \\
\dot{\theta}&=&\frac{1}{g_{\theta \theta}^2}\Big({\cal K}-\frac{l^2}{\sin^2\theta}\Big)\ ,
 \\
\dot{\phi}&=&\frac{l}{g_{\phi \phi}}\ ,
 \\
\dot{t}&=&-\frac{{\cal E}}{g_{tt}}\ ,
\end{eqnarray}
where ${\cal K}$ is the Carter constant corresponding to the total angular momentum.

Restricting the motion of the particle to the plane with $\theta=\rm const$ and $\dot{\theta}=0$ (that is justified by the conservation of the angular momentum), the Carter constant takes the form ${\cal K}=l^2/\sin^2\theta$. Then the equation of the radial motion can be expressed in the form
\begin{eqnarray}
 \dot{r}^2={\cal E}^2-V_{\rm eff}\ ,
\end{eqnarray}
where the effective potential of the motion of neutral particles reads
\begin{eqnarray}\label{effpotentail}
V_{\rm eff} = f(r)\left(1+\frac{l^2}{r^2\sin^2\theta}\right) .
\end{eqnarray}

Now we consider the conditions for the circular motion corresponding to zero radial velocity $\dot{r}=0$) and acceleration  $\ddot{r}=0$. Then one may obtain the radial profiles of the specific angular momentum and specific energy for circular orbits at the equatorial plane ($\theta=\pi/2$) in the following form
\begin{eqnarray}\label{LandE}
l= \frac{r^3 f'(r)}{2f(r)-r f'(r)} \ , \qquad {\cal E}=\frac{2 f(r)^2}{2f(r)-r f'(r)}\ ,
\end{eqnarray}

\begin{figure}[h!]
   \centering
  \includegraphics[width=0.98\linewidth]{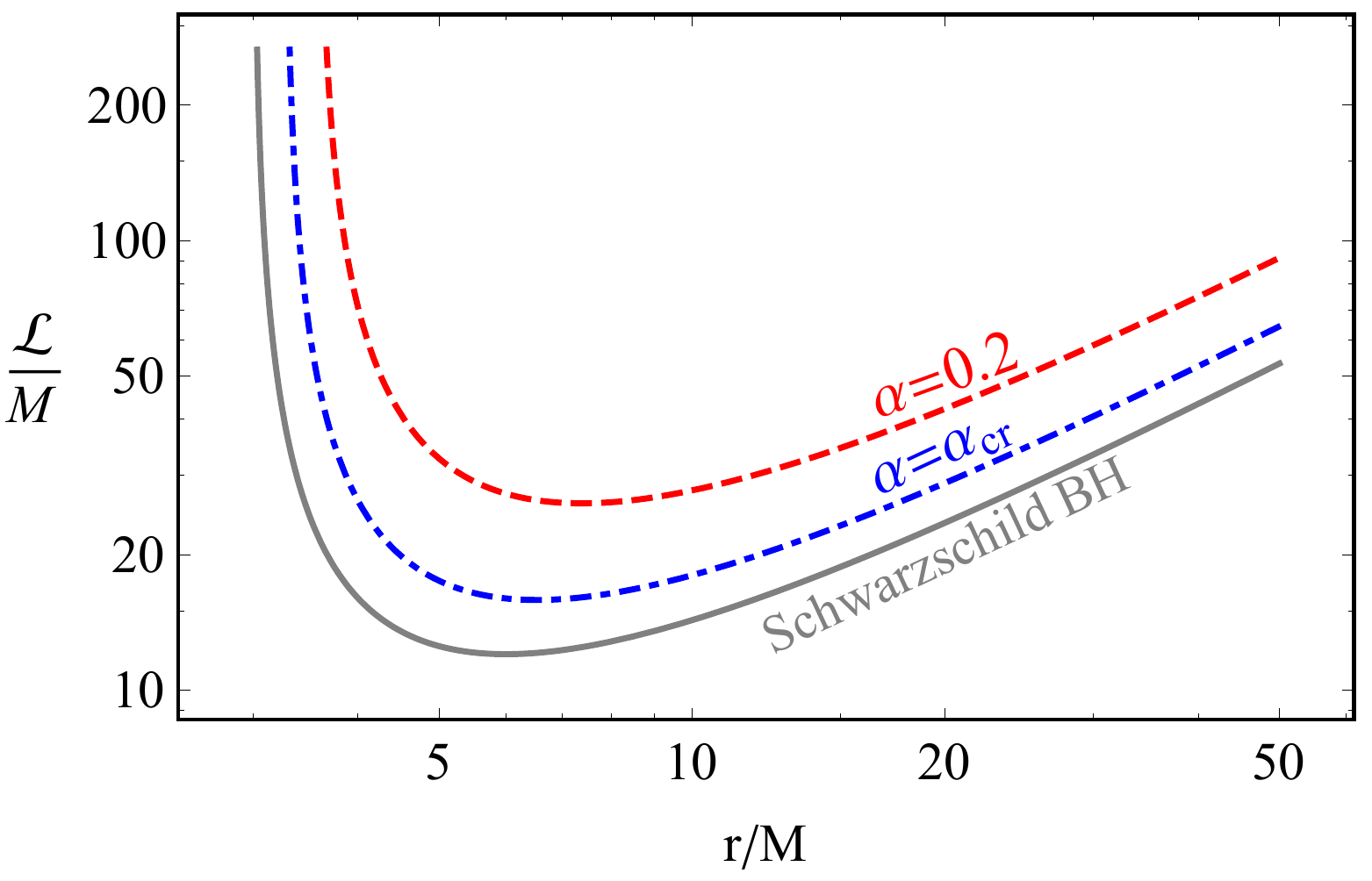}
  \includegraphics[width=0.98\linewidth]{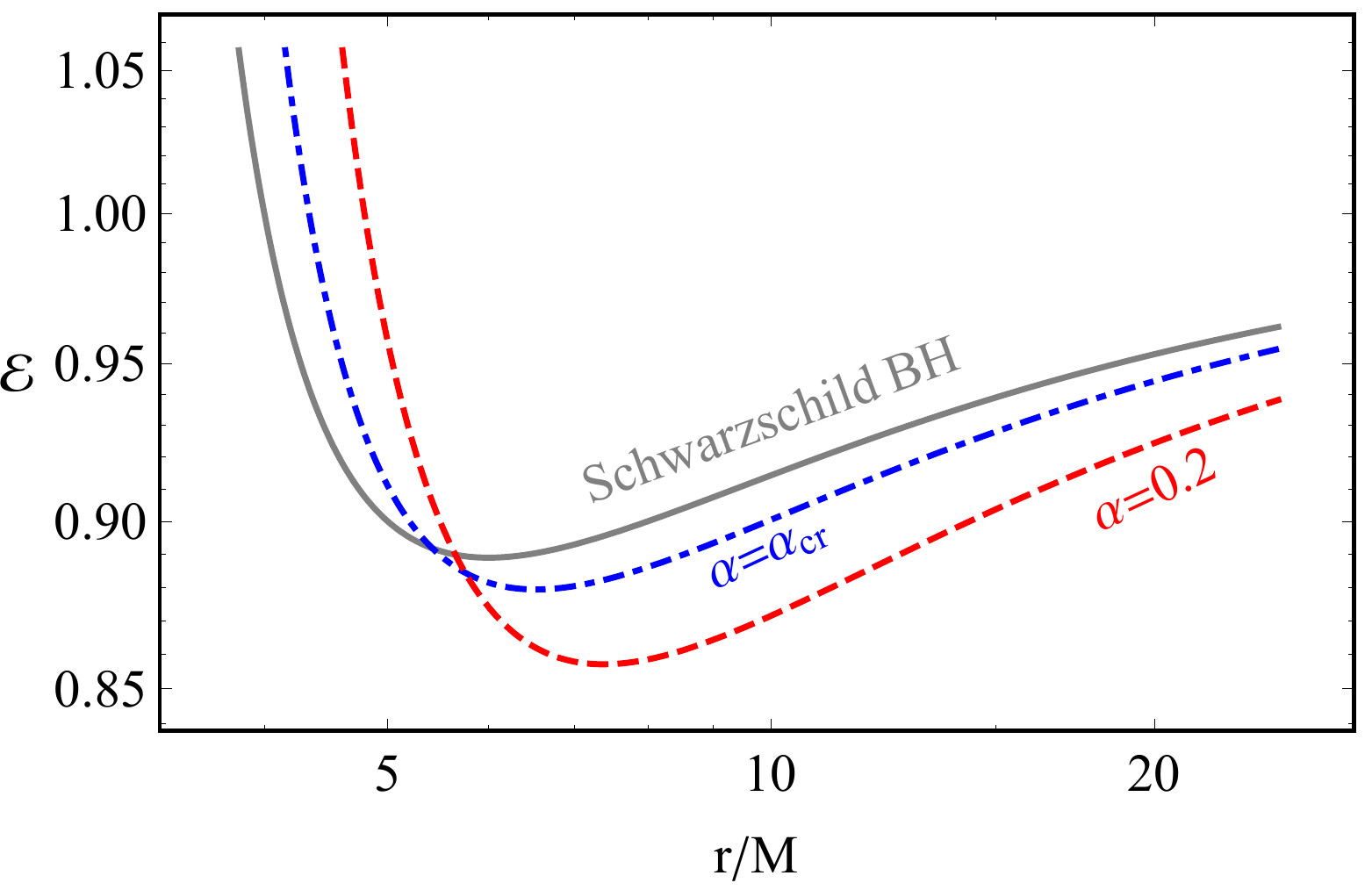}
	\caption{Radial dependence of specific angular momentum (top panel) and energy (bottom panel) for circular orbits  for different values of MOG parameter. \label{landefign}}
\end{figure}

Figure~\ref{landefign} illustrates the radial profiles of the specific energy and angular momentum. One can see that the presence of the MOG parameter causes the increase of the specific angular momentum, the energy decreases. Moreover, the distance corresponding to minimum values of the energy and angular momentum goes outward central black hole with the increase of the MOG parameter.

\subsection{Innermost stable circular orbits}

The innermost stable circular orbits (ISCO) can be defined by the solution of the condition $\partial_{rr}V \geq 0$ with respect to the radial coordinate:
\begin{eqnarray}\label{iscoeq}
f'(r) \left(2 r \frac{f'(r)}{f(r)} -3\right)- r f''(r)\geq 0 \ ,
\end{eqnarray}
where the prime $'$ denotes partial derivative with respect to the radial coordinate. 

It is impossible to find the exact solutions of Eq.(\ref{iscoeq}), however, the limit of the solution at the critical value of MOG parameter can be obtained numerically in the following form 
\begin{equation}\label{iscolimits}
\lim_{\alpha \to \alpha_{cr}}r_{\rm isco}=7.31451\rm M .
\end{equation}

{  It shows that ISCO radius of test particles around Regular MOG black holes with the critic value of the parameter $\alpha$ matches ISCO of the particles in prograte orbits around Kerr black hole with the spin parameter $a/M=0.486402$.} This fact implies to compare the effects of the MOG and spin parameters on ISCO radius and may be used for the estimation of MOG parameter using the observational data for accretion discs.

\subsection{Regular MOG BH versus Kerr BH}
In this subsection we will carry on our detailed analysis on comparison of effects of the spin parameter of rotating Kerr black hole and MOG field on ISCO radius of the test particles.
In theoretical studies and astrophysical measurements of ISCO radius of test particles an indistinguishable problem may appear due to different gravity effects. In most cases black holes are considered as rotational Kerr black holes, however, other type static black holes can provide similar gravitational effects on ISCO radius as rotating black hole. In such cases it is quite difficult to distinguish the difference of gravitational effects of the two different types of black holes. Due to this reason in this subsection we will investigate how the MOG parameter can cover the gravitational effects of spin of Kerr black hole providing the same ISCO radius for test particles mimicking each other and show possible cases of distinguishable values of the MOG and spin parameters which can not cover the gravitational effects of each other. 

The ISCO radius of test particles in retrograde and prograde orbits around Kerr BH can be expressed as~\cite{Bardeen72}
\begin{eqnarray}
r_{\rm isco}= 3 + Z_2 \pm \sqrt{(3- Z_1)(3+ Z_1 +2 Z_2 )} \ ,
\end{eqnarray}
with
\begin{eqnarray} \nonumber
Z_1 &  = & 
1+\left( \sqrt[3]{1+a}+ \sqrt[3]{1-a} \right) 
\sqrt[3]{1-a^2} \ ,
\\ \nonumber
Z_2^2 & = &3 a^2 + Z_1^2 \ .
\end{eqnarray}

\begin{figure}[ht!]
   \centering
  \includegraphics[width=0.98\linewidth]{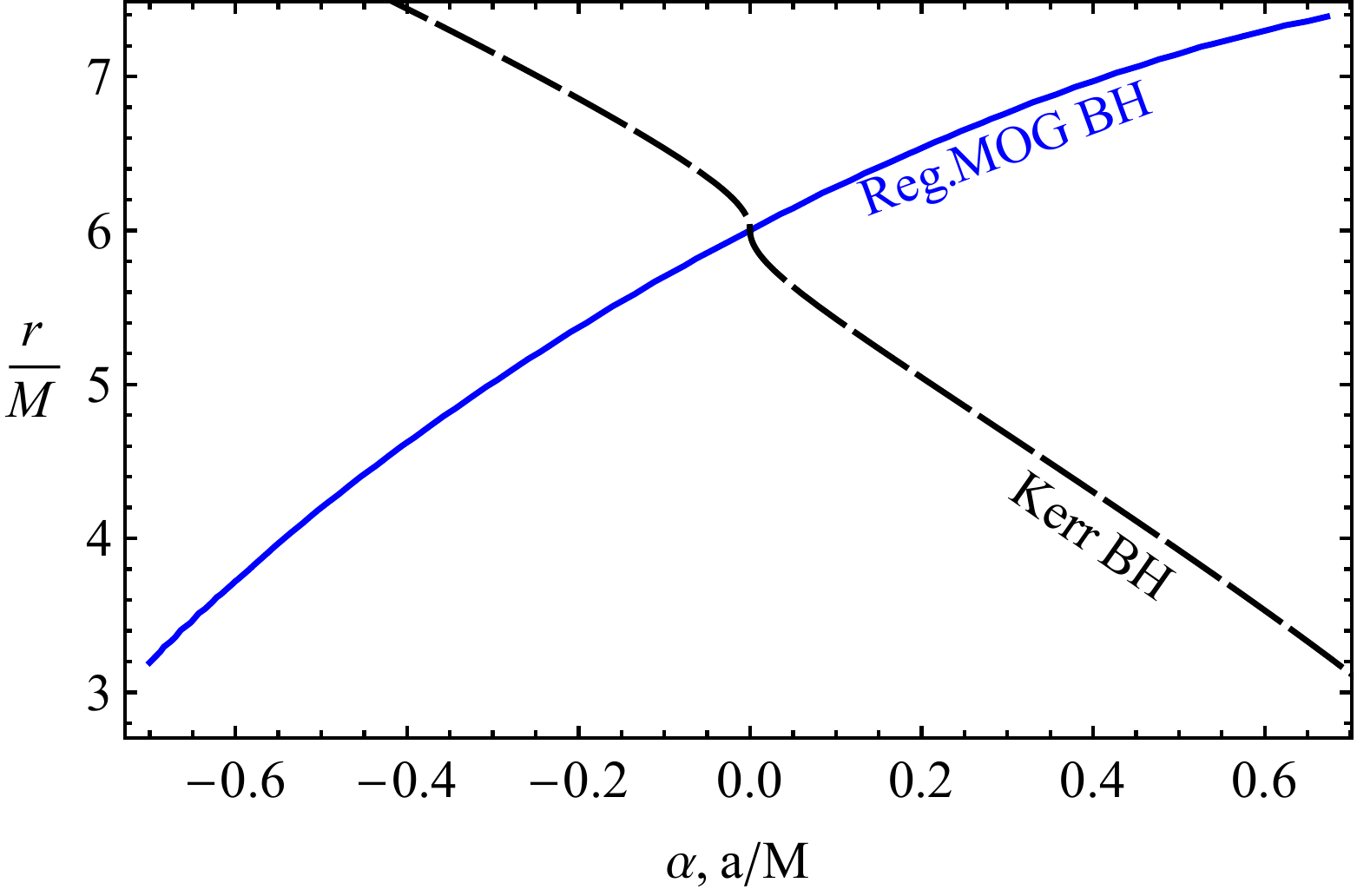}
	\caption{Dependence of ISCO radius on MOG parameter for RMOGBH and on spin parameter for Kerr BH. \label{isco}}
\end{figure} 

Fig.~\ref{isco} shows the dependence  of ISCO radii from MOG parameter and spin parameter for regular MOG black hole and Kerr black hole, respectively. One can see that the negative (positive) values of spin parameter and positive (negative) values of MOG parameter increases the ISCO radii. 

\begin{figure}[h!]
   \centering
  \includegraphics[width=0.9\linewidth]{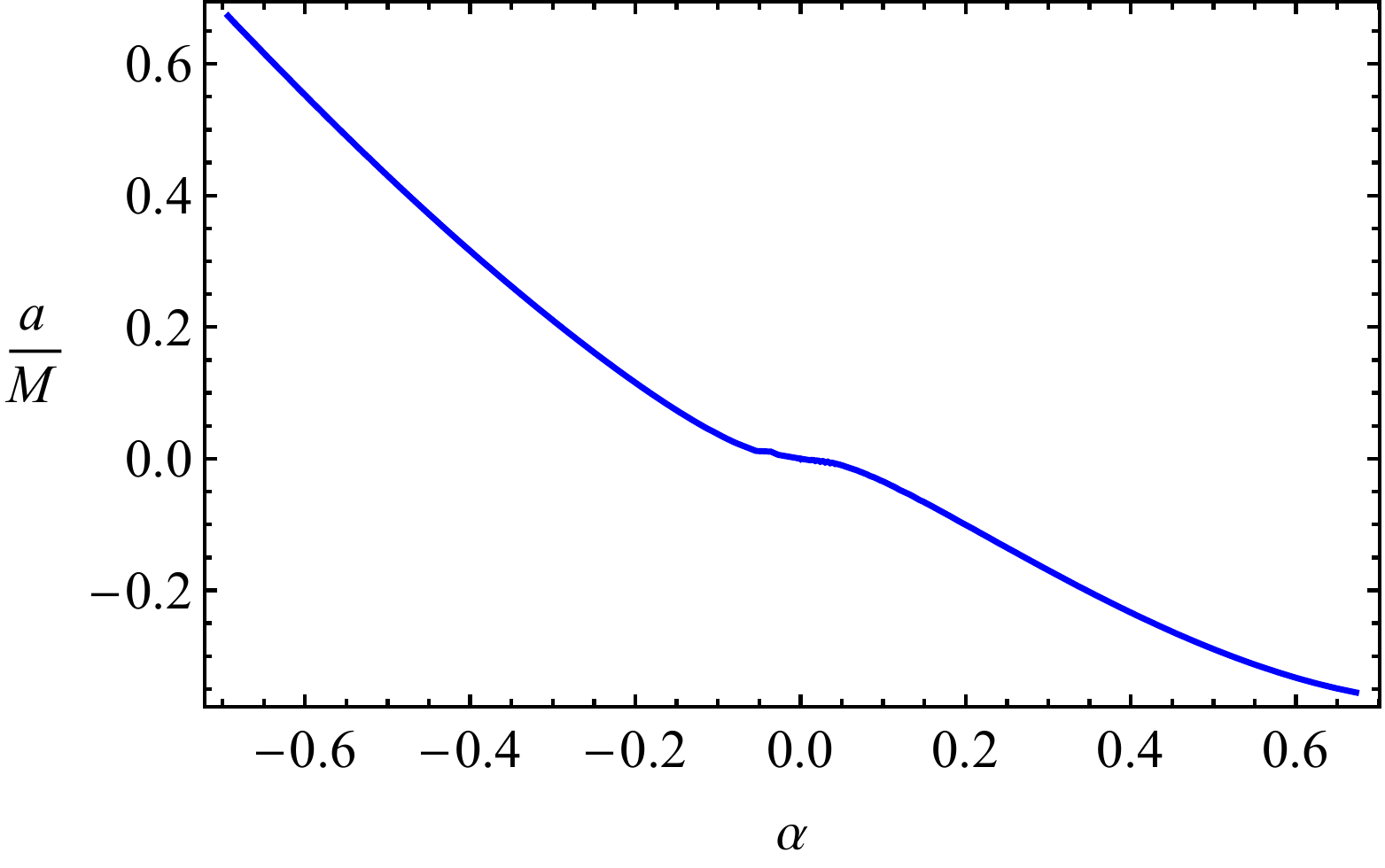}
	\caption{The degeneracy graph between spin and MOG parameters providing the same ISCO radius \label{mimic1}}
\end{figure} 

Fig.~\ref{mimic1} shows the relation between MOG and spin parameters for  the same values of ISCO tadii. It is clear that one cannot fully mimic the spin of Kerr BH by the effects of MOG parameter, since prograde and retrograde motions require different values of MOG parameter. The numerical results show that the MOG parameter can mimic the spin parameter in the range of $a/M \in (-0.4125, \ 0.6946)$ when its values $\alpha \in (-0.6084, \ 0.658)$.

\subsection{The energy extraction efficiency}

Keplerian accretion around an astrophysical black hole is explained by Novikov and Torn as geometrically thin disks~\cite{Novikov73}. The efficiency of energy extraction process in the accretion disk around a black hole is the maximum energy which matter falling in to the central black hole from the disk extracts as radiation energy.  The efficiency of the accretion of the particle can be calculated using the following standard expression
\begin{equation}
\eta=1-{\cal E}\,  \vline_{\,r=r_{\rm ISCO}},
\end{equation}
where ${\cal E}_{\rm ISCO}$ is the energy of the particle at the ISCO which is characterized by the ratio of the binding energy (black hole-particle system) and rest energy of test particle and it can be calculated using the energy of the particles given by Eq.~(\ref{LandE}) at ISCO. 

\begin{figure}[ht!]
   \centering
  \includegraphics[width=0.98\linewidth]{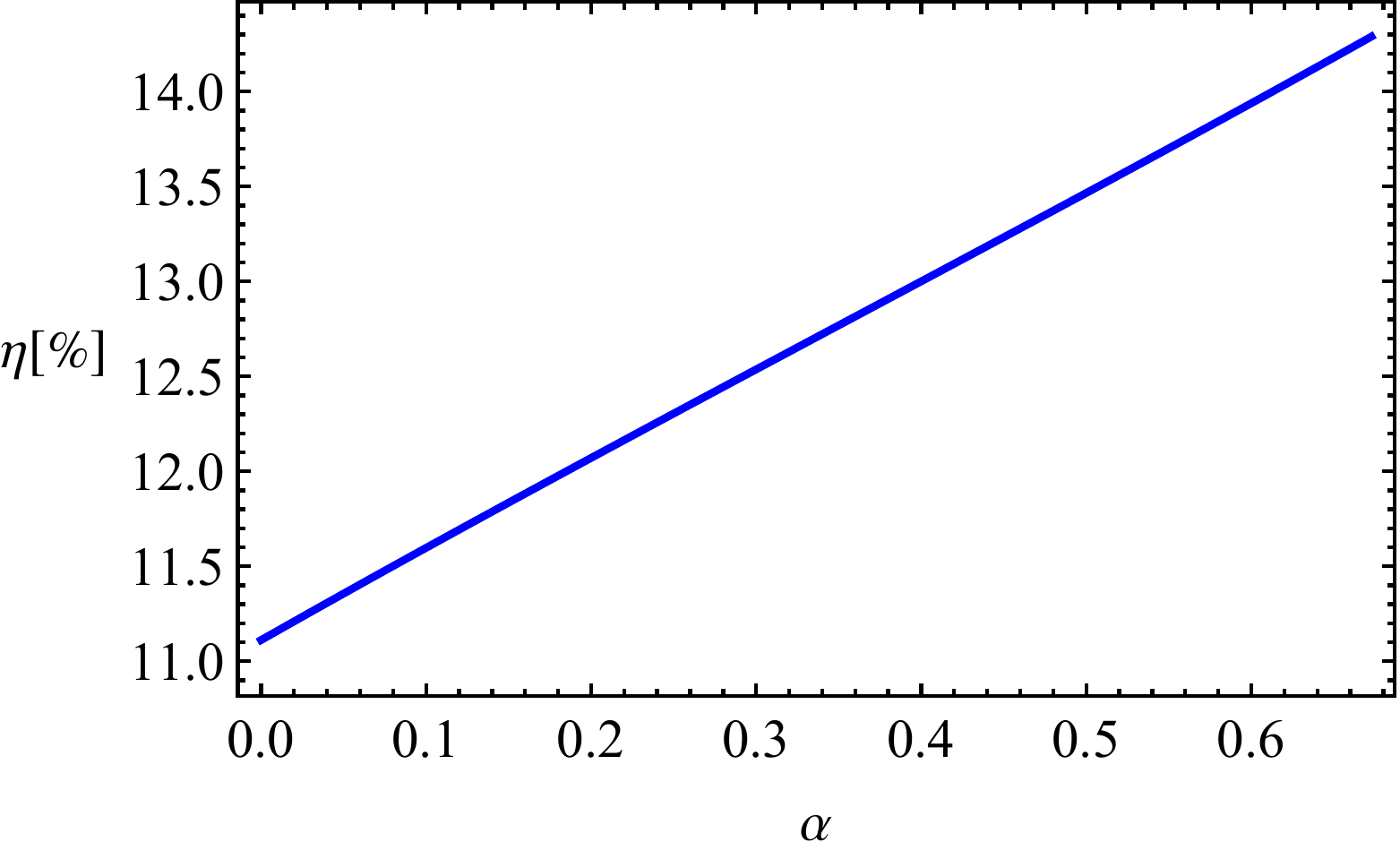}
	\caption{Dependence of energy extraction efficiency on the MOG parameter $\alpha$  \label{efficiency}}
\end{figure}

The effects of MOG parameter on the efficiency of the accretion of the test particle around regular black holes in MOG is presented in Fig.~\ref{efficiency}. One can see that efficiency grows linearly with the increase of  MOG parameter and gives maximum about 15 \% in the range of the parameter $\alpha \in (0,\ \alpha_{\rm cr})$.

\section{Dynamics of charged particles \label{chargedpartmotion}}

In this section we explore the dynamics of electrically charged particles around regular MOG black hole immersed in an external asymptotically uniform magnetic field.  

\subsection{Electromagnetic four-potentials}

In previous subsection~\ref{RRKinvariants} we have shown that at all values of the MOG parameter the spacetime of regular MOG black hole is Ricci flat ($R=0$) outside of outer horizon. So, the Wald method ~\cite{Wald74} is applicable to find the solution of Maxwell equations for the electromagnetic four-potential around regular MOG BH immersed in an external asymptotically uniform magnetic field with the value $B_0$. Using  timelike and spacelike Killing vectors in Ricci flat spacetime one may find the solution for the electromagnetic four-potential in the following form 
\begin{eqnarray}\label{Aft}
A_{\phi} & = & \frac{1}{2}B_0 r^2\sin^2\theta,
\\\nonumber
A_t & = & 0 =A_r=A_{\theta}\ ,
\end{eqnarray}
The non zero components of the electromagnetic field tensor can be found using the definition  $F_{\mu\nu}=A_{\nu,\mu}-A_{\mu,\nu}$ in the following  form
\begin{eqnarray}\label{FFFF}
F_{r \phi}&=&B_0 r\sin^2\theta \ ,
 \\
 F_{\theta \phi}&=&B_0r^2\sin\theta \cos\theta\ .
\end{eqnarray}

One can calculate the non-zero orthonormal components of the external magnetic field around the regular MOG black hole using the following relation
\begin{eqnarray}\label{fields}
B^{\alpha} &=& \frac{1}{2} \eta^{\alpha \beta \sigma \mu} F_{\beta \sigma} u_{\mu}\ ,
\end{eqnarray}
where $\eta_{\alpha \beta \sigma \gamma}$ is the pseudo-tensorial form of the Levi-Civita symbol $\epsilon_{\alpha \beta \sigma \gamma}$ defined by the relations 
\begin{eqnarray}
\eta_{\alpha \beta \sigma \gamma}=\sqrt{-g}\epsilon_{\alpha \beta \sigma \gamma}\, \qquad \eta^{\alpha \beta \sigma \gamma}=-\frac{1}{\sqrt{-g}}\epsilon^{\alpha \beta \sigma \gamma}\ ,
\end{eqnarray}
with $g={\rm det|g_{\mu \nu}|}=-r^4\sin^2\theta$ for spacetime metric (\ref{metric}) and 
\begin{eqnarray}
\epsilon_{\alpha \beta \sigma \gamma}=\begin{cases}
+1\ , \rm for\  even \ permutations
\\
-1\ , \rm for\  odd\  permutations
\\
\, \, 0\ , \, \rm for\ the\ other\ combinations
\end{cases}\ .
\end{eqnarray}
Finally we have
\begin{equation}\label{BrBt1}
    B^{\hat{r}}=B_0 \cos\theta, \\ \qquad B^{\hat{\theta}}=\sqrt{f(r)}B_0\sin \theta\ .
    \end{equation}
 Eq.(\ref{BrBt1}) implies that the orthonormal azimuthal component of the external magnetic field around regular MOG black hole is modified due to the effect of gravitational field.
    
\begin{figure}[h!]
    \centering
  \includegraphics[width=0.97\linewidth]{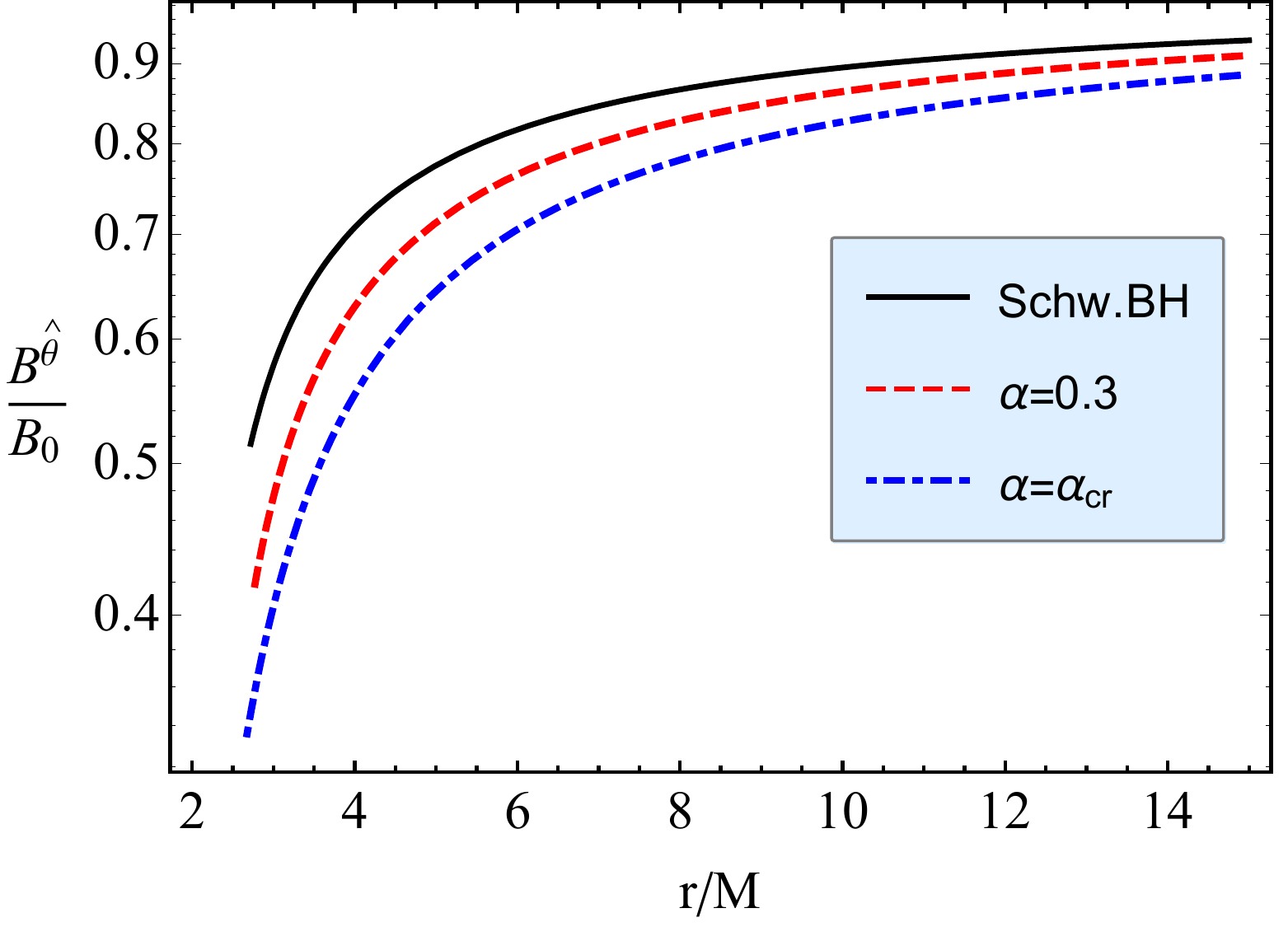}
    \caption{The radial dependence of orthonormal azimuthal component of the external magnetic field $B^{\hat{\theta}}$  with the normalization of its asymptotic value $B_0$.}
    \label{Bt}
\end{figure}

Fig.~\ref{Bt} illustrates the radial dependence of the ortonormal azimuthal component of the magnetic field near the regular MOG black hole for the different values of MOG parameter. One can see from the figure that the azimuthal component of the magnetic field is weakened by effects of the MOG parameter. 

\subsection{Equation of motion}

{Here we will construct the equation of motion for electrically charged particles with electric charge $q$, around the regular MOG black hole in the external magnetic field using the Hamilton-Jacobi equation}
\begin{eqnarray}\label{HJ}
g^{\mu \nu}\Big(\frac{\partial {\cal S}}{\partial x^{\mu}}-q A_{\mu}\Big)\Big(\frac{\partial {\cal S}}{\partial x^{\nu}}-q A_{\nu}\Big)=-m^2\ .
\end{eqnarray}
Since, $t$ and $\phi$ are the Killing variables, the action can be written in the following form
\begin{eqnarray}\label{action}
{\cal S}=-{\cal{E}} t+L \phi+{\cal S}_r(r)+{\cal S}_{\theta}(\theta) \ .
\end{eqnarray}

Four velocities of the charged particle in the equatorial plane where $\theta=\pi/2, \ \dot{\theta}=0$ can be expressed in the separable form as follows:
\begin{eqnarray}\label{eqmotionch}
\nonumber
\dot{t}&=&\frac{{\cal E}}{f(r)}\ ,
\\
\nonumber
\dot{r}^2&=&{\cal E}^2-f(r)\Big[1+\Big(\frac{l}{r}-\omega_{\rm B} r\Big)^2\Big]\ ,
\\
\dot{\phi}&=&\frac{l}{r^2}-\omega_{\rm B}\ ,
\end{eqnarray}
where $\omega_{\rm B}=qB_0/(2mc)$ is the cyclotron frequency related to interaction of magnetic field and charged particle
\begin{eqnarray}
\dot{r}^2={\cal E}^2-V_{\rm eff} \ .
\end{eqnarray}

Effective potential for the particle can be easily found substituting Eq.(\ref{action}) into Eq.(\ref{HJ}) in the following simple form
\begin{equation}
    V_{\rm eff}= f(r)\left[1+\left(\frac{l}{r \sin \theta}-\omega_B r \sin\theta\right)^2\right] \ .
\end{equation}

\subsection{ISCO}
 
We use the following standard conditions to study of stable circular orbits in the equatorial plane
\begin{eqnarray}\label{iscocond}
V_{\text{eff}}={\cal E}\ ,
\qquad
V'_{\text{eff}}=0\ ,
\qquad
V''_{\text{eff}} \geq 0\ .
\end{eqnarray}
For the solutions of the equation $V'_{\text{eff}}=0$ (with respect to $r$) to be the radii of the circular orbits for the particle with specfic charge $q$, the specific angular momentum of the particle must be subject to following relation:
 \begin{eqnarray}\label{lpmeq}
 l_{\pm} &= & \frac{r}{{\cal Z}r-2}\left[ 2{\cal Z}r+r^2 \left(4\omega_{\rm B}^2-{\cal Z}^2\right)\pm \omega_{\rm B}{\cal Z} \right]\ ,
 \\\label{epmeq}
 \nonumber
 {\cal E}_{\pm} &=& \frac{2f(r)}{{\cal Z}r-2}\Big\{2-{\cal Z}r+4\omega^2_{\rm B}r^2 
 \\ &\pm &  \omega_{\rm B} r \left[2{\cal Z}r+r^2 \left(4\omega_{\rm B}^2-{\cal Z}^2\right)\right]\Big\}\ .
 \end{eqnarray}
 where ${\cal Z}=\partial \ln f(r) / \partial r$\ .
 
 Eqs. (\ref{lpmeq}) and (\ref{epmeq}) show that the specific angular momentum and energy of the charged particles $l$ and ${\cal E}$ are symmetric under replacement of $\omega_{\rm B} \to -\omega_{\rm B}$, i.e.  $(l,{\cal E})_{-}(\omega_{\rm B}>0)=(l,{\cal E})_{+}(\omega_{\rm B}<0)$.
 
 \begin{figure}[ht!]
   \centering
  \includegraphics[width=0.98\linewidth]{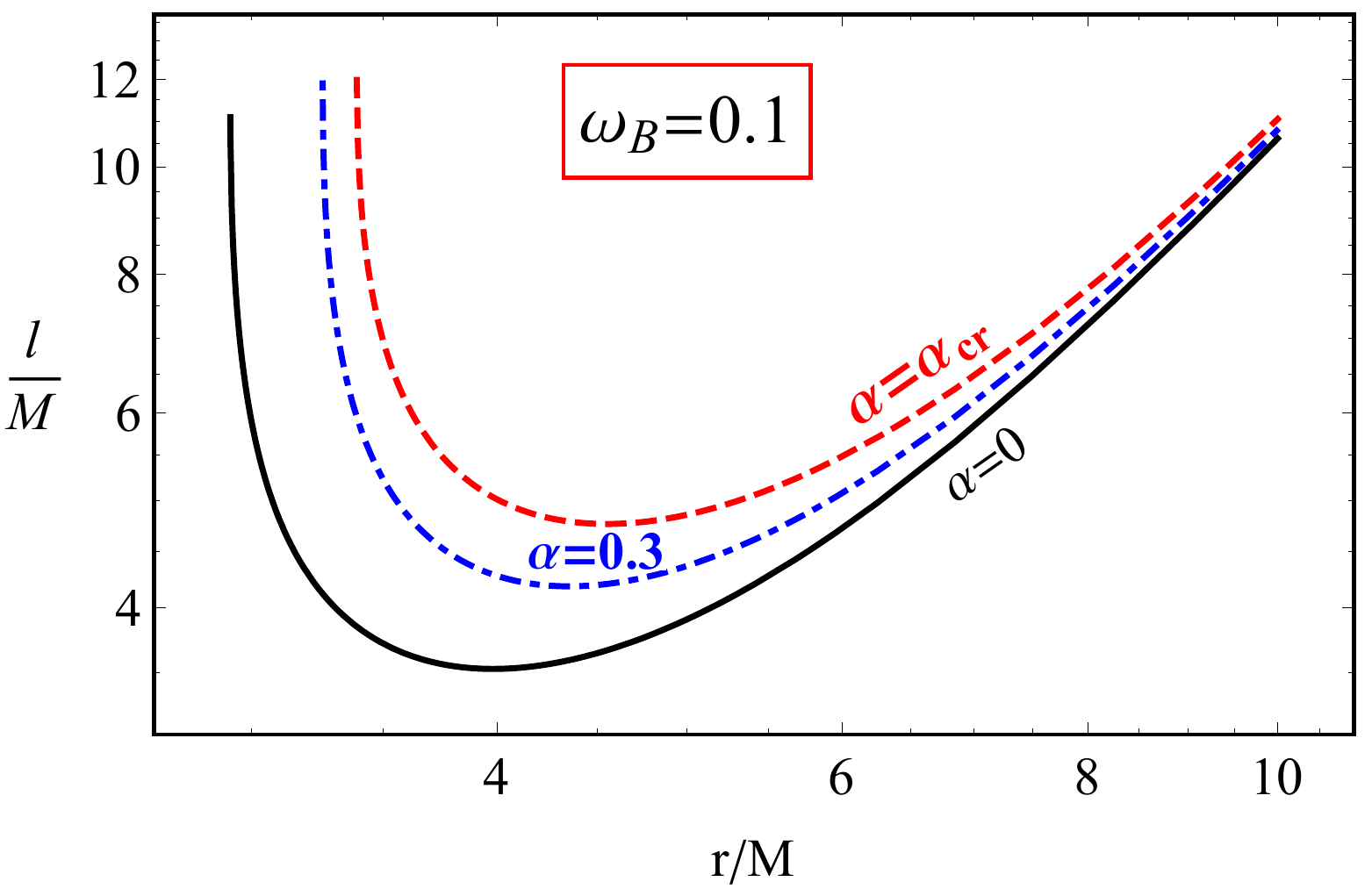}
  \includegraphics[width=0.98\linewidth]{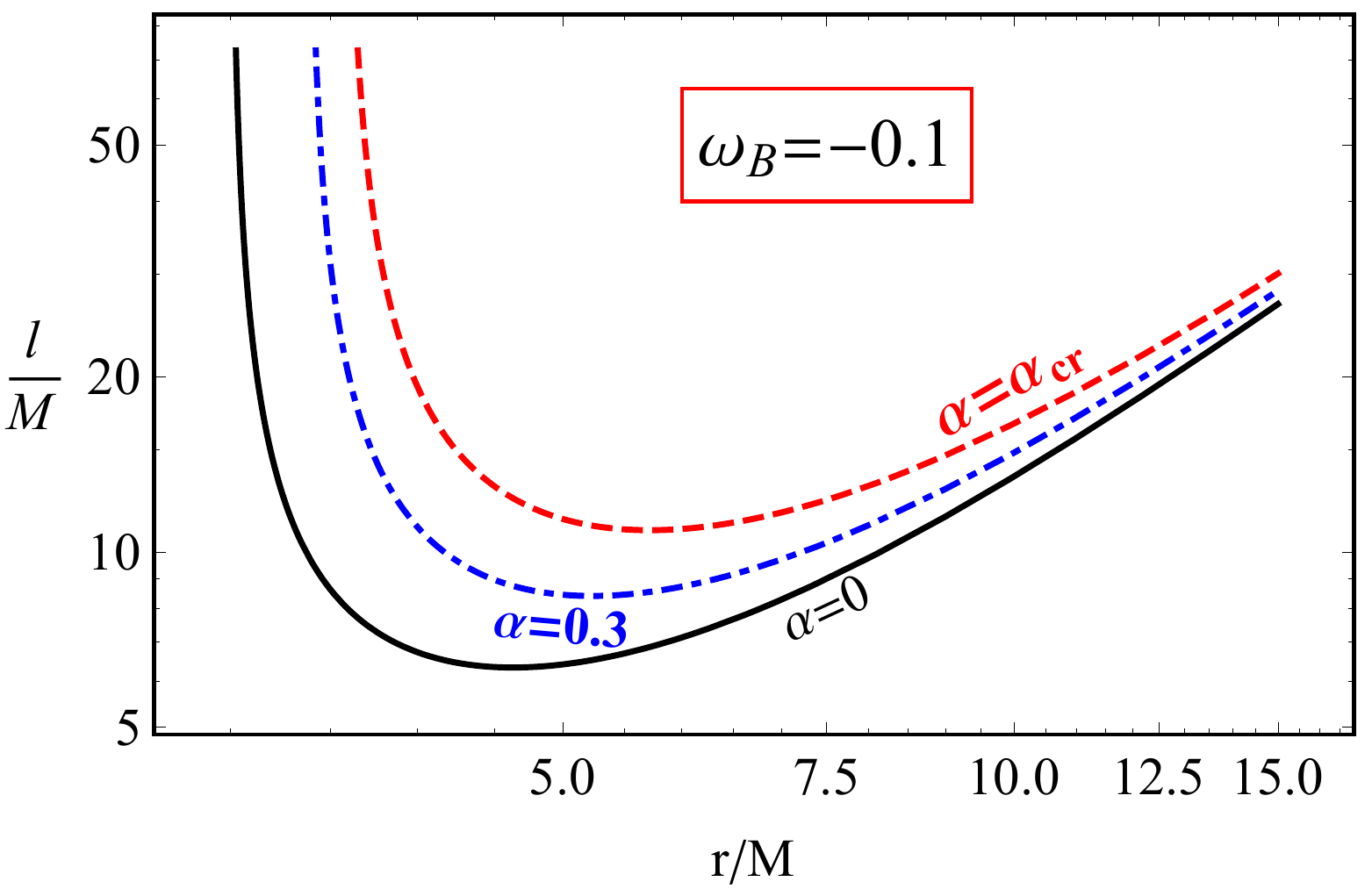}
	\caption{Radial dependence of specific angular momentum of positively (top panel) and negatively (bottom panel) charged particles for circular orbits for different values of $\alpha$. \label{lw}}
\end{figure}
 
 Figure~\ref{lw} illustrates the dependence of specific angular momentum of positive (on the top panel) and negative (on the bottom panel) charged particles from radial coordinate for the different values of the MOG parameter. One can see that the increase of MOG parameter causes the increase of the angular momentum and decrease of the minimum distance for circular orbits. Moreover, the distance where the angular momentum is minimum shifts outward the central black hole. From the comparison of the top and bottom parameters that the minimum values of the specific angular momentum of the charged particles corresponding to circular orbits smaller for positive charged particle than negative one. 
 
 Now we will analyse the effects of MOG parameter on specific energy of the charged particle in the orbits of circular motion by plotting Eq.(\ref{epmeq}).
 
 \begin{figure}[ht!]
   \centering
  \includegraphics[width=0.98\linewidth]{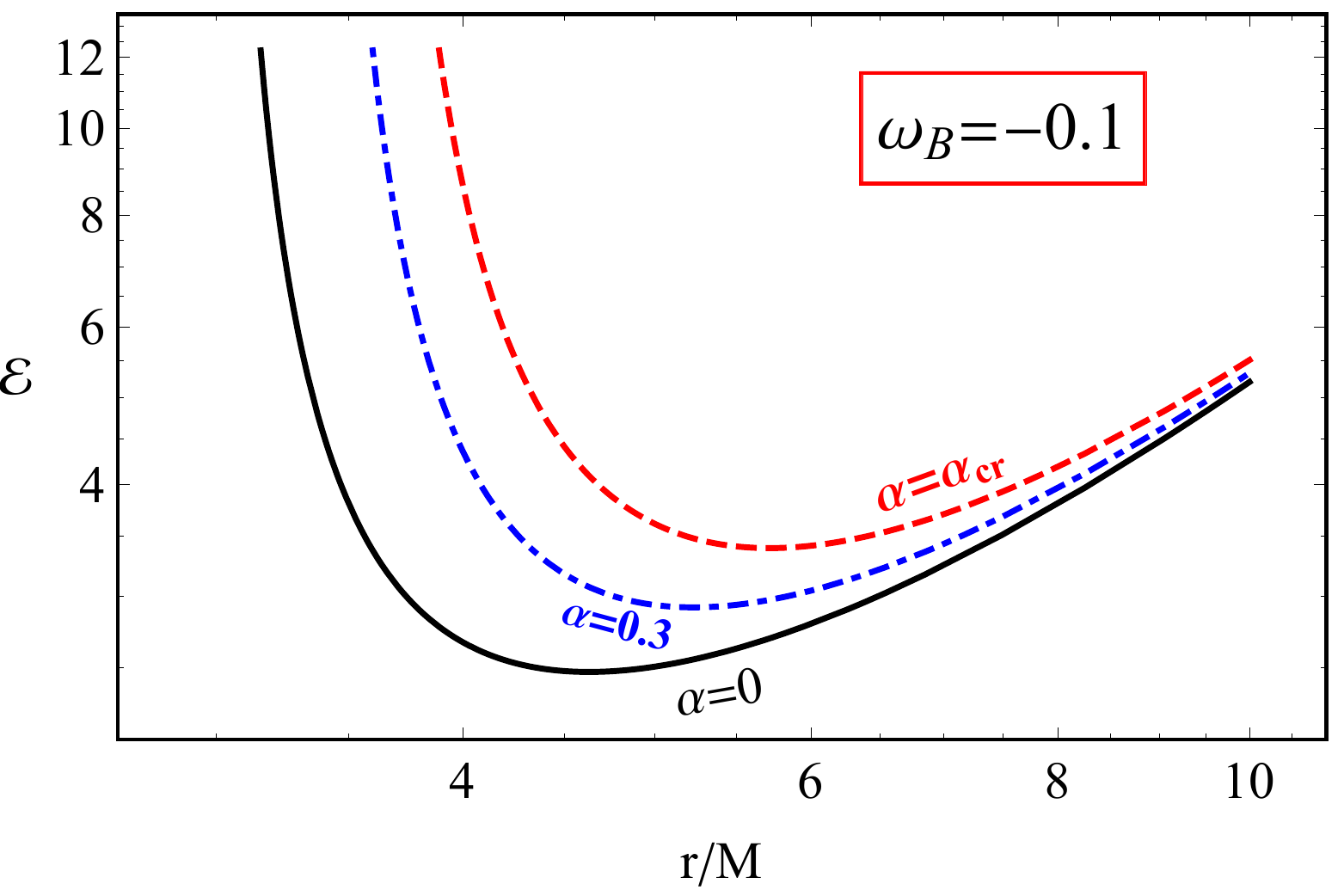}
  \includegraphics[width=0.98\linewidth]{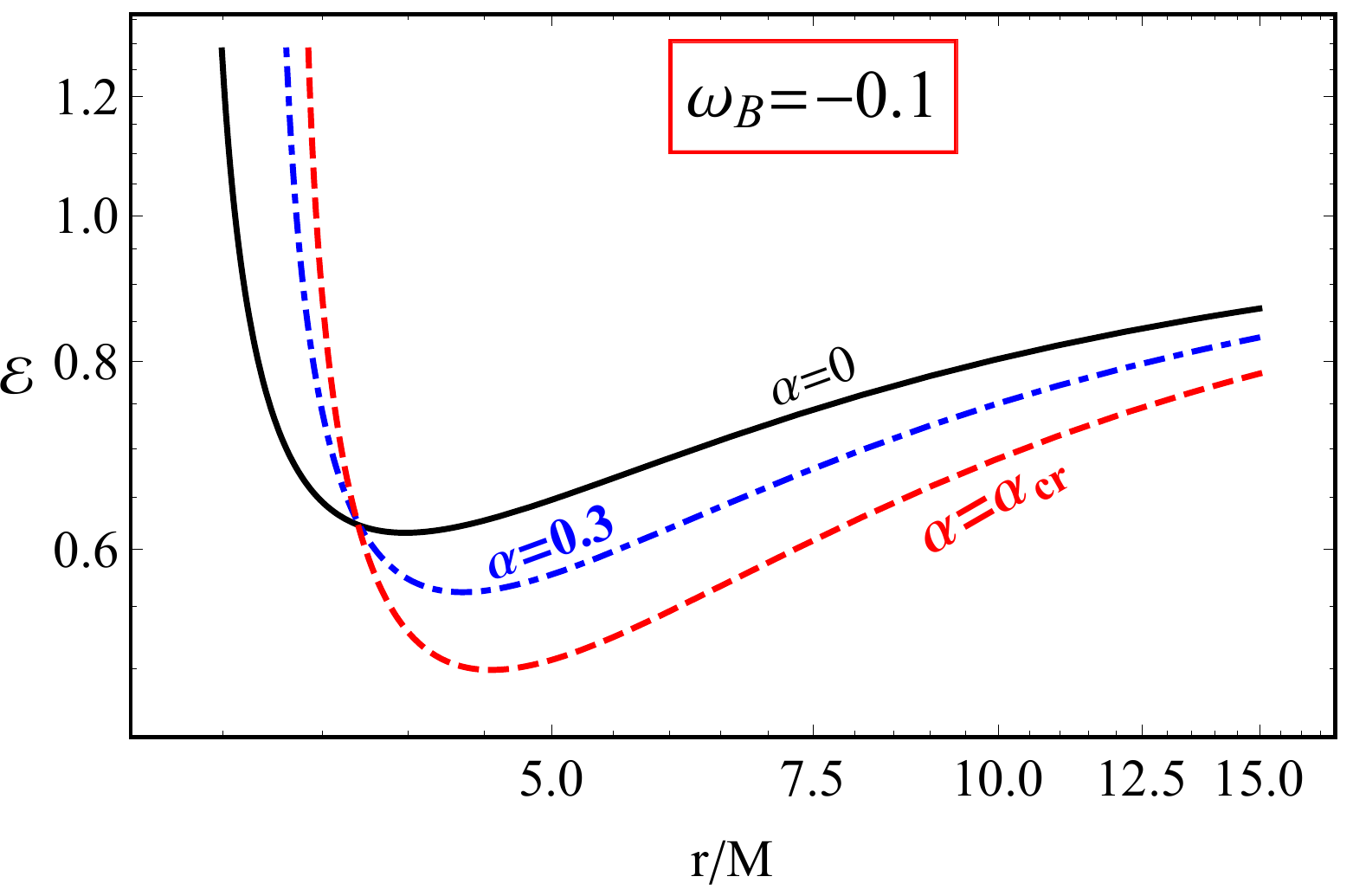}
	\caption{Radial dependence of specific energy of positively (top panel) and negatively charged particles (bottom panel) for circular orbits for different values of $\alpha$. \label{Ew}}
\end{figure}

Fig.~\ref{Ew} illustrates the radial dependence of specific energy of oppositely charged particles for different values of MOG parameter.

 One may derive the ISCO equation from the last part of the condition for stable circular orbits as follows:
 \begin{eqnarray} \label{iscoweq}
 \nonumber
 &&2 f(r) \Big\{2 r^2 \omega _B \left[\omega _B (2-2 r \mathcal{Z} (r \mathcal{Z}-1))+\mathcal{Q} \mathcal{Z} (3-2 r \mathcal{Z})\right]\\ \nonumber
 &&+\mathcal{Q}^2 (3-2 r \mathcal{Z})\Big\}+r^2 f''(r) \Big\{\Big[(r \mathcal{Z}-2)^2
\\
&&\pm 4 r \omega _B (r \mathcal{Z}-1) \left(\mathcal{Q} \pm r \omega _B (r \mathcal{Z}-1)\right)\Big]+\mathcal{Q}^2\Big\}\geq 0\ .
\end{eqnarray}
where $\mathcal{Q}=2{\cal Z}r+r^2 \left(4\omega_{\rm B}^2-{\cal Z}^2\right).$

\begin{figure}[h!]
   \centering
  \includegraphics[width=0.98\linewidth]{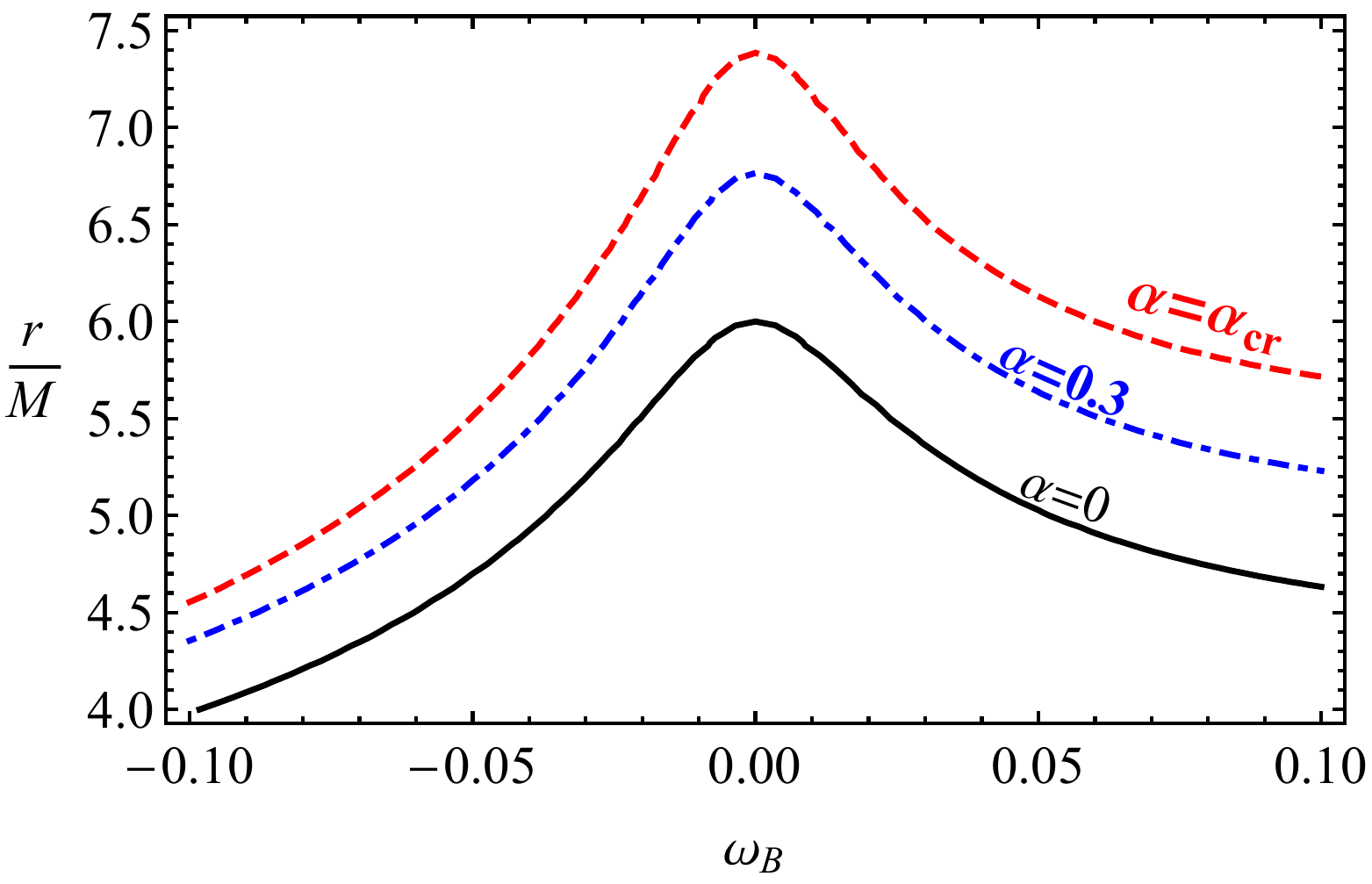}
	\caption{Dependence of ISCO radius from cyclotron frequency $\omega_{\rm B}$ for different values of MOG parameter $\alpha$. \label{iscow}}
\end{figure}

The ISCO radius of the charged particles is described by the root of Eq.(\ref{iscoweq}) with respect to the radial coordinate. However, it is impossible to solve analytically the equation and see the effects of the MOG parameter on ISCO radius of the test charged particle. So, we solve it numerically and plot the relation between ISCO radius and cyclotron frequency for different values of MOG parameter in Fig.~\ref{iscow}.
One can see that the increase of both positive and negative values of the magnetic interaction parameter causes the decrease of the ISCO radius, while ISCO radius growths with the increase of the MOG parameter. Moreover, the ISCO radius is bigger for charged particles with positive magnetic interaction parameters than the particles with negative ones due to different feature of Lorentz interaction.  

\section{The motion of magnetized particles \label{magnetpartmotion}}

The magnetized particle motion may be considered one of possible ways to test a theory of gravity and corresponding spacetime metric around black hole. Magnetized neutron stars observed as radio pulsars with the precisely measured periods of rotation can be approximated as test magnetized particles in the close environment of Srg A* in the center of Galaxy. No pulsar has been observed in the central parsec of SgrA* so far mainly due to the scattering of radio waves by dense, turbulent and ionized plasma in the SMBH close environment rather than due to an intrinsic absence of radio pulsars in the vicinity to the supermassive black hole. However, the recent success of Event Horizon Telescope in imaging the central SMBH of M87 \cite{2019ApJEHT} might lead to advances in observation of stellar objects in vicinity of Sgr A*.

To describe the motion of magnetized particles in the presence of the external asymptotically uniform magnetic field around the BH we use following Hamilton-Jacobi equation~\cite{deFelice}
\begin{eqnarray}\label{H-J}
g^{\mu \nu}\frac{\partial {\cal S}}{\partial x^{\mu}} \frac{\partial {\cal S}}{\partial x^{\nu}}=- m_{\rm eff}^2\ ,
\end{eqnarray}
where 
\begin{equation}
   m_{\rm eff}=m-\frac{1}{2} {\cal D}^{\mu \nu}F_{\mu \nu}
\end{equation}
{and ${\cal D}^{\mu \nu}F_{\mu \nu}$ is the term responsible for the interaction of magnetized particle with external magnetic field. We will assume that the particle is a magnetic dipole.} The polarization tensor ${\cal D}^{\alpha \beta}$ must satisfy the following condition
\begin{eqnarray}\label{dexp}
{\cal D}^{\alpha \beta}=\eta^{\alpha \beta \sigma \nu}u_{\sigma}\mu_{\nu}\ , \qquad {\cal D}^{\alpha \beta }u_{\beta}=0\ ,
\end{eqnarray} 
where $\mu^{\nu}$ is the four-vector of magnetic dipole moment of the magnetized particle. The interaction term of the Hamilton-Jacobi equation (\ref{HJ}) can be determined using the electromagnetic field tensor $F_{\alpha \beta}$ which can be expressed in terms of electric $E_{\alpha}$ and magnetic $B^{\alpha}$ fields as
\begin{eqnarray}\label{fexp}
F_{\alpha \beta}=u_{\alpha}E_{\beta}-u_{\beta}E_{\alpha}-\eta_{\alpha \beta \sigma \gamma}u^{\sigma}B^{\gamma}\ . 
\end{eqnarray}
Taking into account the condition given in Eq.(\ref{dexp})  we have 
\begin{eqnarray}\label{DF1}
{\cal D}^{\alpha \beta}F_{\alpha \beta}=2\mu^{\hat{\alpha}}B_{\hat{\alpha}}=2\mu B_0 {\cal L}[\lambda_{\hat{\alpha}}]\ ,
\end{eqnarray}
where $\mu =\sqrt{|\mu_{\hat{i}}\mu^{\hat{i}}|}$ is the norm of the magnetic dipole moment of the particle and ${\cal L}[\lambda_{\hat{\alpha}}]$ is a function of the effects of the frame of reference comoving with the magnetized particle rotating around the BH. That function depends on the spatial coordinates, as well as other parameters and defines the tetrads $\lambda_{\hat{\alpha}}$ attached to the fiducial comoving observer~\cite{deFelice}.

As we mentioned above our aim is to study the effects of Regular MOG BHs on the magnetized particle and for simplicity we assume the magnetic interaction between the magnetized particle and external magnetic field is weak enough (due to either weakness of the external magnetic field or the small magnetic moment of the particle), so we can use the approximation of $\left({\cal D}^{\mu \nu}F_{\mu \nu} \right)^2 \to 0$. Moreover, we also assume the direction of the magnetic dipole moment of the magnetized particle to be perpendicular to the equatorial plane, with the components  $\mu^{i}=(0,\mu^{\theta},0)$. 

Since spacetime symmetries are conserved in the presence of the external asymptotically uniform magnetic field, still we have two integrals of motion (i.e. conserved quantities) for the magnetized particle: angular momentum $p_{\phi}= L$ and energy $p_t = -E$ of the magnetized particle, respectively. The expression for the action of magnetized particles that allows to separate variables in the Hamilton-Jacobi equation (\ref{H-J}) can be written as
\begin{eqnarray}\label{action}
{\cal S}=-E t+L\phi +{\cal S}_r(r)\ .
\end{eqnarray}

The expression for radial motion of magnetized particles at the equatorial plane where $\theta=\pi/2$ ($p_{\theta}=0$) can be found by substituting Eq.(\ref{DF1}) to Eq.(\ref{HJ}) taking account of Eq.(\ref{action}) in the following form: 
\begin{eqnarray}
\dot{r}^2={\cal{E}}^2-V_{\rm eff}(r;l,\beta,\alpha)\ . 
\end{eqnarray}
The analytic expression of the effective potential takes the form
\begin{eqnarray}\label{effpot}
V_{\rm eff}(r;l,\beta,\alpha)=f(r)\left(1+\frac{l^2}{r^2}-\beta {\cal L}[\lambda_{\hat{\alpha}}]\right)\ ,
\end{eqnarray}
where $\beta = 2\mu B_0/m$ is the magnetic coupling parameter which corresponds to the interaction term ${\cal D}^{\mu \nu}F_{\mu \nu}$ in the Hamilton-Jacobi equation (\ref{HJ}). For the typical neutron star orbiting around supermassive black hole with magnetic dipole moment $\mu=(1/2)B_{\rm NS}R^3_{\rm NS}$ the magnetic parameter is 
\begin{eqnarray}\label{betaNS}
\beta \simeq \frac{1}{250}\left(\frac{B_{\rm NS }}{10^{12} \rm G}\right)\left(\frac{R_{\rm NS}}{10^6 \rm cm}\right)^3\left(\frac{B_{\rm ext}}{10\rm G}\right)\left(\frac{m_{\rm NS}}{M_{\odot}}\right)^{-1} \ . 
\end{eqnarray}
We recall the conditions for circular orbits
\begin{eqnarray} \label{conditions}
\dot{r}=0 \ , \qquad \partial_r V_{\rm eff}=0\ .
\end{eqnarray}
The first equation of (\ref{conditions}) together with Eq.~(\ref{effpot})
gives the following expression 
\begin{eqnarray}\label{betafunc1}
\beta(r;l,{\cal{E}},\alpha)=\frac{1}{ {\cal L}[\lambda_{\hat{\alpha}}]}\Bigg(1+\frac{l^2}{r^2}-\frac{{\cal{E}}^2}{f(r)}\Bigg)\ .
\end{eqnarray}
{The explicit expressions for the orthonormal tetrad carried the fiducial observer ${\cal L}[\lambda_{\hat{\alpha}}]$ can be formulated for circular motion at the equatorial plane around spherical symmetric black hole in the following form
\begin{eqnarray}
\lambda_{\hat{t}}&=&e^{\Psi}\left( \partial_t+\Omega \partial_{\phi} \right)\ , \\\nonumber
\lambda_{\hat{r}}&=&e^{\Psi}\left[-\frac{\Omega r}{\sqrt{f(r)}}\partial_t-\frac{\sqrt{f(r)}}{r}\partial_{\phi} \right]\sin(\Omega_{FW} t)\\&+&\sqrt{f(r)}\cos (\Omega_{FW}t)\partial_r\ , \\
\lambda_{\hat{\theta}}&=&\frac{1}{r}\partial_{\theta}\ , \\\nonumber
\lambda_{\hat{\phi}}&=&e^{\Psi}\left[\frac{\Omega r}{\sqrt{f(r)}}\partial_t+\frac{\sqrt{f(r)}}{r}\partial_{\phi} \right]\cos(\Omega_{FW}t)\\&+&\sqrt{f(r)}\sin(\Omega_{FW}t)\partial_r\ ,
\end{eqnarray} 
where $\Omega_{FW}$ is Fermi-Walker angular velocity \cite{deFelice} and
\begin{equation}
    \label{epsi}
 e^{-2\Psi}=f(r)-\Omega^2 r^2 \ ,
\end{equation}
with $\Omega$ is the angular velocity of the particles measured by a distant observer defined as
\begin{eqnarray}
\Omega=\frac{d\phi}{d t}=\frac{d\phi/d\tau}{d t/d\tau}=\frac{f(r)}{r^2}\frac{l}{{\cal{E}}}\ .
\end{eqnarray}
We will study the motion of a magnetized particle orbiting at the equatorial plane assuming the magnetic dipole moment of the magnetized particle is always perpendicular to the equatorial plane and parallel to the external magnetic field. The orthonormal components of the external magnetic field measured by the observer comoving with the magnetized particle take the following form
\begin{eqnarray}\label{Bcomp}
B_{\hat{r}}=B_{\hat{\phi}}=0\ , \qquad B_{\hat{\theta}}=B_0f\,e^{\Psi}\ .
\end{eqnarray}
The induced electric field measured by the comoving observer can be written as
\begin{eqnarray}
E_{\hat{r}}&=&B_0\Omega r\cos(\Omega_{FW}t)\sqrt{f(r)}e^{\Psi}\ , \\
E_{\hat{\theta}}&=&0 \ , \\
E_{\hat{\phi}}&=&B_0\Omega r \sin(\Omega_{FW}t)\sqrt{f(r)}e^{\Psi}\ .
\end{eqnarray}
In case when the Fermi-Walker and the particles angular velocities measured by proper observer are zero ($\Omega_{FW}=\Omega=0$), the above tetrad turns to tetrad of the proper observe (see \cite{Rezzolla01} in the case $a=0$) and the magnetic field components in Eq.(\ref{Bcomp}) equals to the components in Eq.(\ref{BrBt1}) and the induced electric field vanishes.  
}

The interaction term in Eq.~(\ref{H-J}) can be found using Eqs.(\ref{Bcomp}) and (\ref{DF1}) in the form
\begin{eqnarray}\label{DF2}
{\cal D}^{\mu \nu}F_{\mu \nu}=2\mu B_0f(r)\,e^{\Psi}\ , \end{eqnarray}
which can be used to find the characteristic function 
\begin{eqnarray}\label{lambda}
{\cal L}[\lambda_{\hat{\alpha}}]=e^{\Psi}\, f(r)\ .
\end{eqnarray}
Finally, we will find the exact form of the magnetic coupling parameter $\beta(r;l,{\cal E},\alpha)$ using Eqs. (\ref{lambda}), (\ref{epsi}), and (\ref{betafunc1}) in the following form: 
\begin{eqnarray}\label{betafinal}
  \beta(r;l,{\cal E},\alpha)=\left(\frac{1}{f(r)}-\frac{l^2}{{\cal E}^2 r^2}\right)^{\frac{1}{2}}\Bigg(1+\frac{l^2}{r^2}-\frac{{\cal E}^2}{f(r)}\Bigg).\ \
\end{eqnarray}

Eq.(\ref{betafinal}) indicates that magnetized particle with given specific energy ${\cal E}$ and angular momentum $l$ can be at the circular orbit at a given radial coordinate $r$.

\begin{figure}[ht!]
  \centering
   \includegraphics[width=0.9\linewidth]{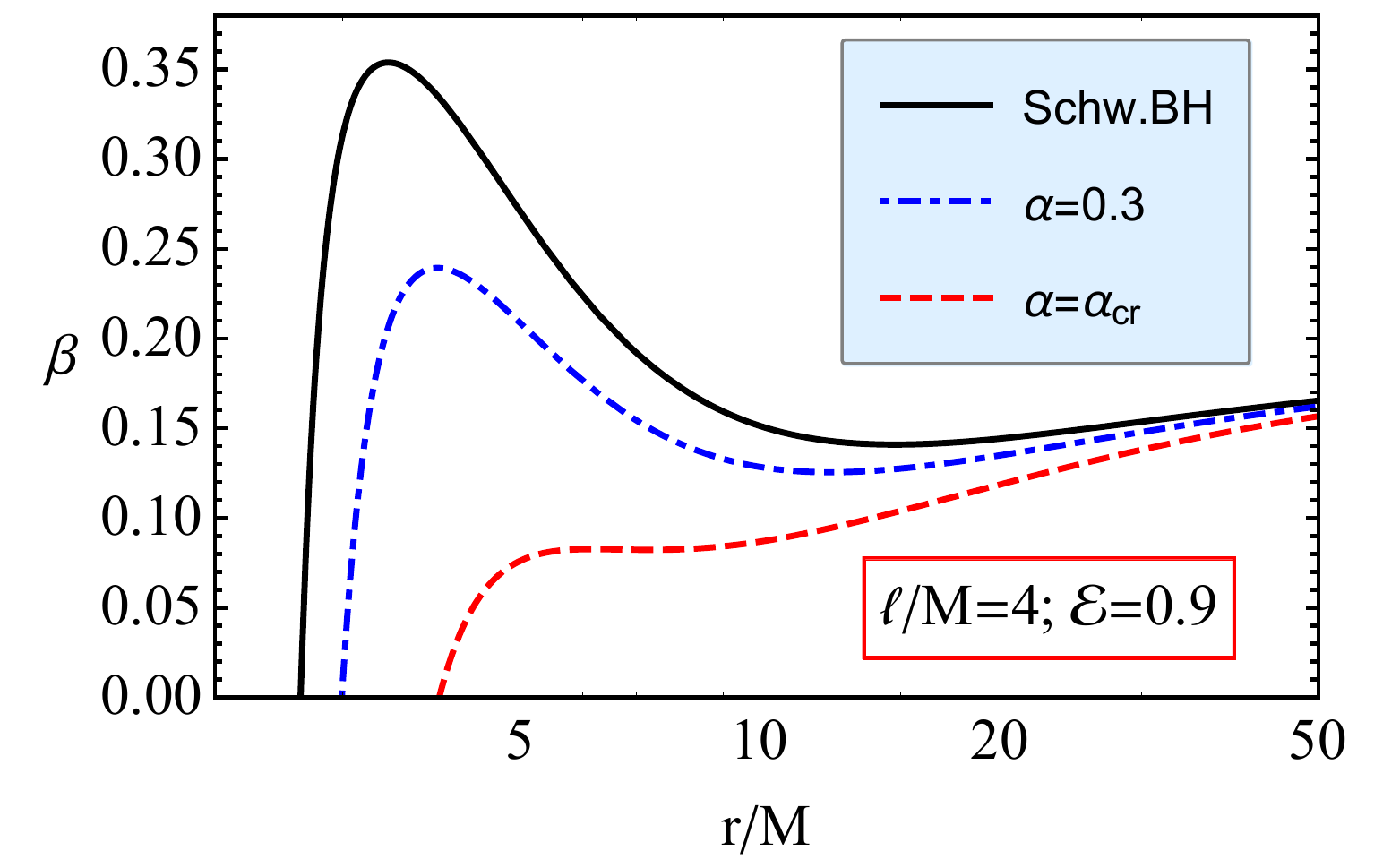}
   \caption{Radial profiles of the function of magnetic interaction parameter $\beta$ for the different values of the parameters $\alpha$. We have used the following values of the specific angular momentum $l/M=4$ and the energy of the particle ${\cal E}=0.9$. \label{betafig}}  
\end{figure}
  
The magnetic coupling/interaction  parameter $\beta$ as a function of radial coordinate for the different values of $\alpha$ parameter is demonstrated in Figure~\ref{betafig}. One may see that {the extrema of the function merge at the critical value of MOG parameter and the function becomes non-decreasing. At the values of MOG parameter lesser than the critical one we have different "magnetic zones" around the black hole according to the increase and decrease of coupling parameter, while at the critical value there's no zoning.}

Now we will concentrate on the studies of the stable circular orbits for magnetized particles questioning in which values of the magnetic coupling parameters the particle with a given values of specific energy and angular momentum could be in a circular orbits and it can be redefined in terms of $\beta$ using the following set of conditions:
\begin{eqnarray}\label{conditionstab}
\beta =\beta(r;l,{\cal E},\alpha), \qquad \partial_r \beta(r;l,{\cal E},\alpha)=0\ .
\end{eqnarray}
Eq.(\ref{conditionstab}) is system of equations and now we interested in the solution of the system of equations with five variables $\beta,r,l,{\cal E}$ and MOG parameter $\alpha$ given in Eq.(\ref{conditionstab}) can be obtained in terms of any two of the five independent variables.

The expressions for minimum value for the specific energy ${\cal E}$ of a magnetized particle at stable circular orbit with the minimum value of the magnetic coupling parameter, can be found by solving the second part of Eq.~(\ref{conditionstab}) with respect to the energy ${\cal E}$ and using the coupling parameter $\beta$ together with radial coordinate $r$ as free parameters:
\begin{eqnarray}\label{emin}
&&{\cal E}_{\rm min}(r;l,\alpha) = \frac{l \left(\alpha  (\alpha +1) M^2+r^2\right)^{3/2}}{\sqrt{M(\alpha +1)} r^2}\\\nonumber
&& \times \Bigg[1+ \frac{(\alpha +1) M r^2 \left(\alpha  M-2 \sqrt{\alpha  (\alpha +1) M^2+r^2}\right)}{\left(\alpha  (\alpha +1) M^2+r^2\right)^2}\Bigg]\\\nonumber && \times \Bigg\{\alpha ^2 (\alpha +1) M^3-\alpha  M r^2 +r^2 \sqrt{\alpha  (\alpha +1) M^2+r^2} \\\nonumber && -2 \alpha  (\alpha +1) M^2 \sqrt{\alpha  (\alpha +1) M^2+r^2}\Bigg\}^{-\frac{1}{2}}
\end{eqnarray}

When $\alpha \to 0$ the Eq.~(\ref{emin}) turns the energy of magnetized particle in the case of pure Schwarzschild BH which has the form (see \cite{deFelice})
\begin{eqnarray}
{\cal E}_{\rm min}(r;l)=\frac{l}{\sqrt{Mr}}\left(1-\frac{2M}{r}\right)\ .
\end{eqnarray}

\begin{figure}[ht!]
  \centering
   \includegraphics[width=0.95\linewidth]{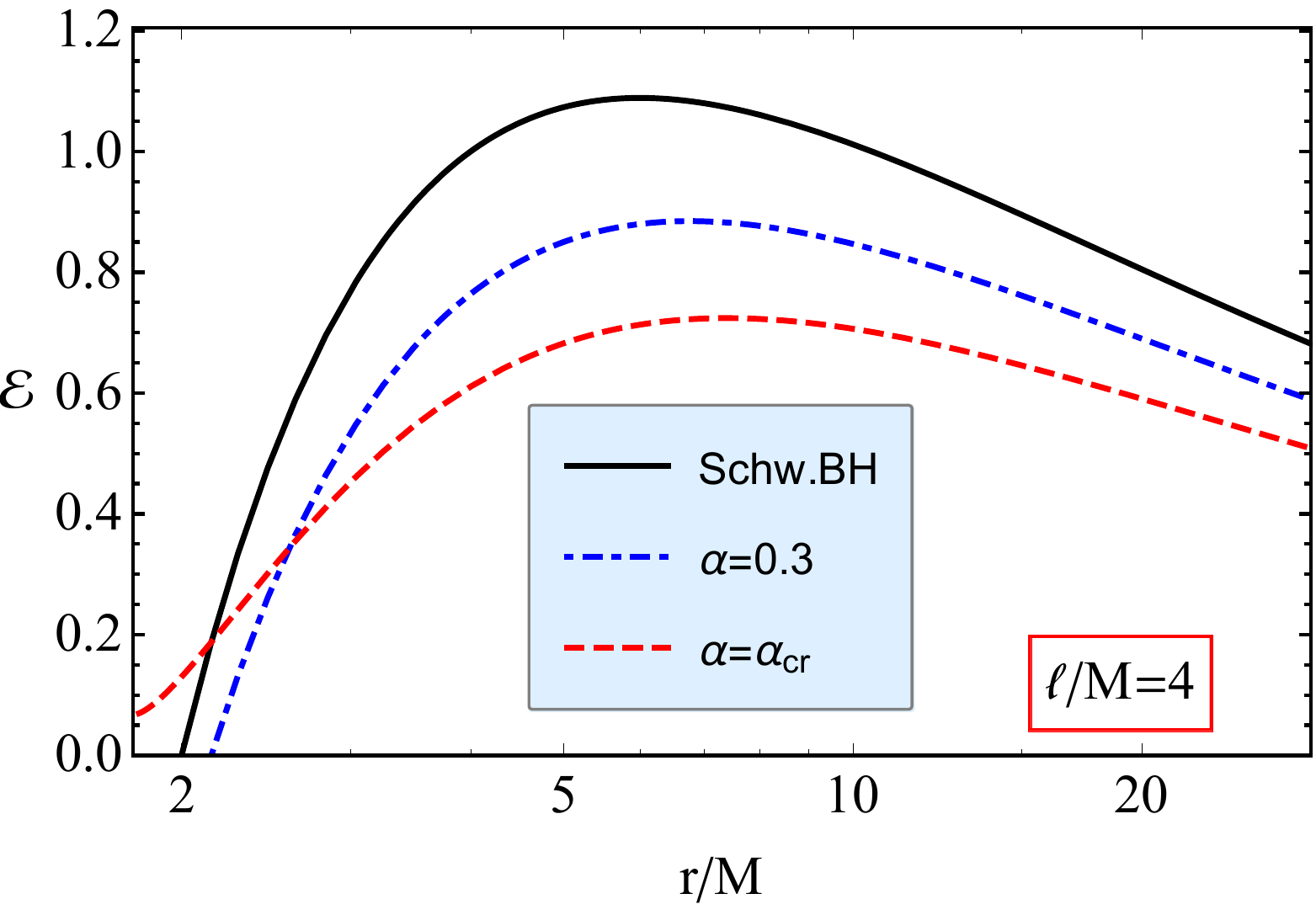}
      \caption{Radial dependence of the specific energy of magnetized particles for minimize the magnetic coupling parameter for the different values of the MOG parameter, at $l/M=4$ \label{eminfig}}
\end{figure}

Fig.~\ref{eminfig} presents radial dependence of the minimum values of specific energy of the magnetized particles allowing them to be in circular orbits for the different values of the parameter $\alpha$ for the fixed values of the specific angular momentum $l/M=4$. One can see from the figure that there is a maximum in the curve of the radial profiles of the energy which allows the magnetic coupling parameter to be minimum. The maximum value of the specific energy decreases under the effects of increase of the MOG parameter while the distance where the energy is maximum increases with the increase of the parameter $\alpha$.

The expression for minimum of magnetic interaction parameter $\beta$ can be obtained by substituting Eq.~(\ref{emin}) into Eq.~(\ref{betafinal}) in the following form 
\begin{widetext}
\begin{eqnarray}\label{betamineq}
\nonumber
&&\beta_{\rm min}(r;l,\alpha) = \sqrt{\alpha ^3 (\alpha +1)^3 M^6+3 \alpha ^2 (\alpha +1)^2 M^4 r^2+(\alpha +1) M r^4 \left(5 \alpha  M-3 \sqrt{\alpha  (\alpha +1) M^2+r^2}\right)+r^6}\\
&& \times  \frac{\sqrt{\alpha  (\alpha +1) M^2+r^2}}{\alpha ^2 (\alpha +1)^2 M^4+(\alpha +1) M r^2 \left(3 \alpha  M-2 \sqrt{\alpha  (\alpha +1) M^2+r^2}\right)+r^4} \\\nonumber && \times \Bigg\{ 1-\frac{l^2 \left(\alpha ^3 (\alpha +1)^3 M^6+3 \alpha ^2 (\alpha +1)^2 M^4 r^2+(\alpha +1) M r^4 \left(5 \alpha  M-3 \sqrt{\alpha  (\alpha +1) M^2+r^2}\right)+r^6\right)}{(\alpha +1) M r^4 \left(\alpha ^2 (\alpha +1) M^3-2 \alpha  (\alpha +1) M^2 \sqrt{\alpha  (\alpha +1) M^2+r^2}+r^2 \sqrt{\alpha  (\alpha +1) M^2+r^2}-\alpha  M r^2\right)} \Bigg\}
\end{eqnarray}
\end{widetext}

Since it is not clear from the expression (\ref{betamineq}) the effects of the MOG parameter on the extreme value of the magnetic coupling parameter, we will analysis it by plotting 
\begin{figure}[ht!]
  \centering
   \includegraphics[width=0.98\linewidth]{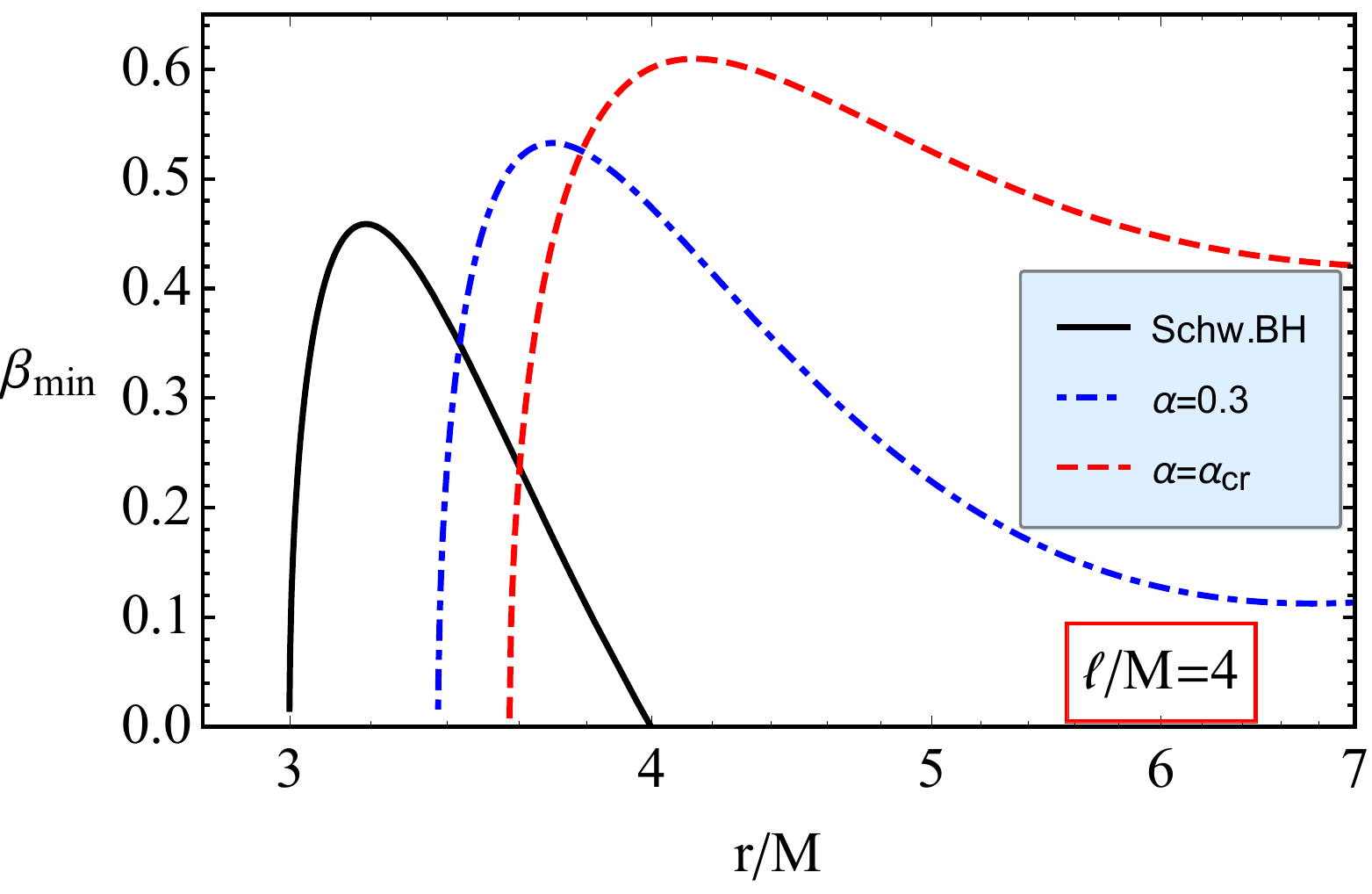}
    \caption{Radial profiles of minimum magnetic interaction parameter--$\beta_{\rm min}$ for the different values of the parameter $\alpha$. We use the value of the specific angular momentum $l=4M$. \label{betaminfig}}
\end{figure}

Fig.~\ref{betaminfig} represents minimal magnetic interaction parameter as a function of radial coordinate for different values of the parameter $\alpha$. One can see from this figure that there is also a maximum in the relation. {It is evident that the maximum values of $\beta_{\rm{min}}$ grow with the increase of MOG parameter as well as the size of the area where such minimum does not exist.} Moreover, the distance where $\beta_{\rm min}$ has maximum shifts outward from the central object as the MOG parameter increases.

In fact that the upper limit for stable circular orbits corresponds to minimum value of specific angular momentum and the extreme of the magnetic coupling parameter corresponds to the minimum angular momentum. One can find the extreme values of the magnetic coupling parameter by solving the equation $\partial_r \beta_{\rm min}(r;l,\alpha)=0$ with respect to $l$ and we have the following quite longer form of expression for it
\begin{widetext}
\begin{eqnarray}\label{lmineq}
\nonumber
l_{\rm min}(r; \alpha)&=&\frac{(\alpha +1) M r^3}{\left[\alpha  (\alpha +1) M^2+r^2\right]^3} \left\{\sqrt{\alpha  (\alpha +1) M^2+r^2}-\alpha  M\right\} \left[r^2-\alpha  (\alpha +1) M^2\right] \\\nonumber &\times & \Bigg\{2+ \frac{(\alpha +1) M r^2 }{\left[\alpha  (\alpha +1) M^2+r^2\right]^3}  \Bigg[\alpha ^2 (\alpha +1) M^3+\left(2 \alpha  (\alpha +1) M^2+r^2\right) 
 \left(\alpha  M-3 \sqrt{\alpha  (\alpha +1) M^2+r^2}\right)\Bigg]\Bigg\}^{-\frac{1}{2}} \\ & \times & \Bigg[1+ \frac{(\alpha +1) M r^4 }{\left[\alpha  (\alpha +1) M^2+r^2\right]^3}  \Bigg(2 \alpha  M-3 \sqrt{\alpha  (\alpha +1) M^2+r^2}\Bigg)\Bigg]^{-\frac{1}{2}}
\end{eqnarray}
\end{widetext}

\begin{figure}[h!]
  \centering
   \includegraphics[width=0.98\linewidth]{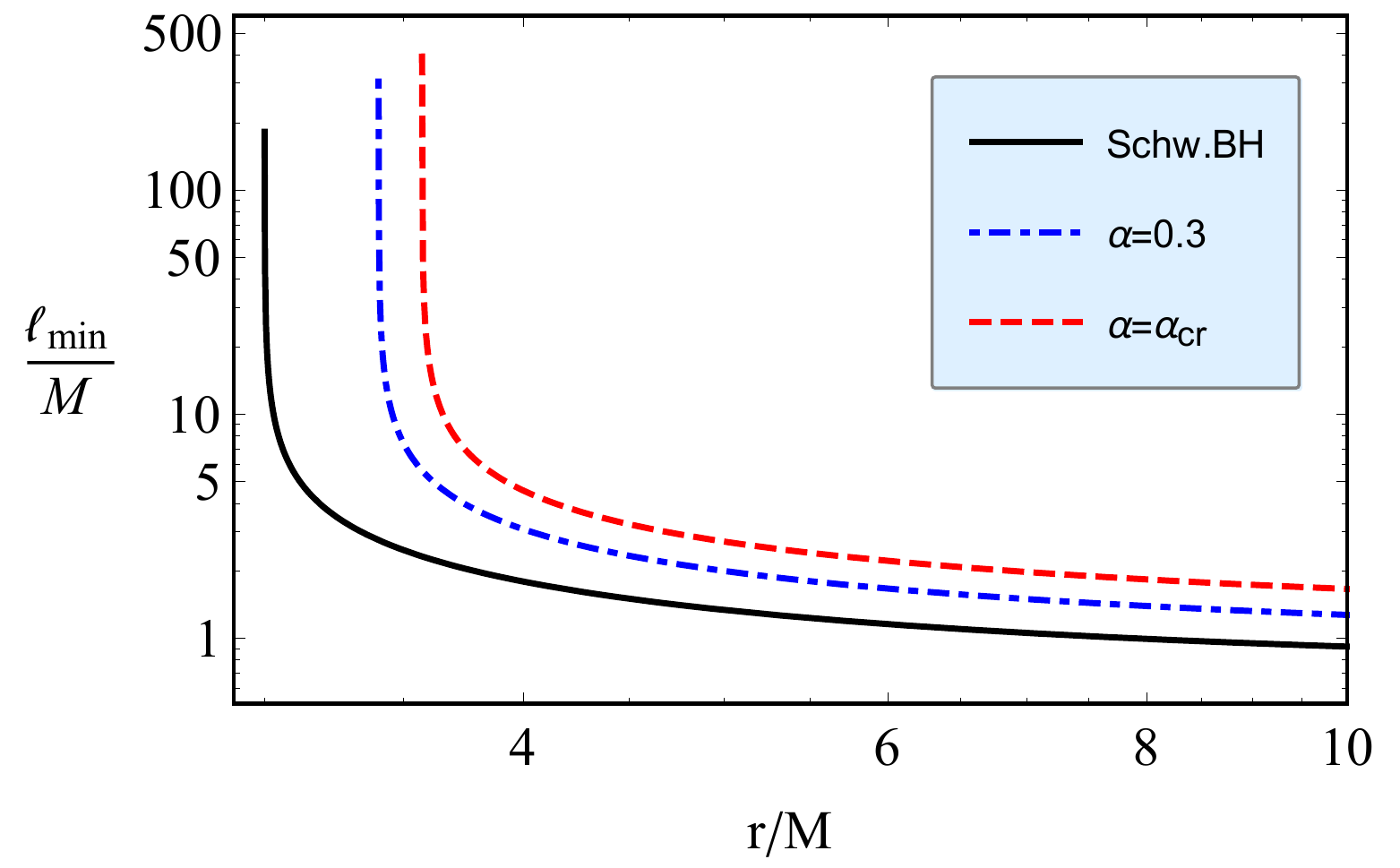}
    \caption{Radial profiles of minimal specific angular momentum, $l_{\rm min}$, for different values of the MOG parameter $\alpha$. \label{lminfig}}
\end{figure}

Fig.~\ref{lminfig} illustrates the minimum value of the specific angular momentum of magnetized particles corresponding to circular stable motion as a function of radial coordinate for different values of the parameter MOG $\alpha$. One may see from the figure that the maximum of the angular momentum increases with increasing the parameter of MOG and the distance where it is maximum shifts outward from the central black hole.    

One may find the extreme magnetic coupling parameter inserting Eq.(\ref{lmineq}) in to Eq.~(\ref{betamineq}) in the following form 

\begin{eqnarray}\nonumber
&&\beta_{\rm extr}(r;\alpha)=\Bigg\{1+\frac{(\alpha +1) M r^4}{\left[\alpha  (\alpha +1) M^2+r^2\right]^3}\\\nonumber && \times \Big(2 \alpha  M -3 \sqrt{\alpha  (\alpha +1) M^2+r^2}\Big)\Bigg\}^{-\frac{1}{2}} \\\nonumber && \times \Bigg\{1+\frac{(\alpha +1) M r^2}{2 \left[\alpha  (\alpha +1) M^2+r^2\right]^3} \\\nonumber && \times \Big[\alpha ^2 (\alpha +1) M^3+\left[3 \alpha  (\alpha +1) M^2+r^2\right]\Big] \\ && \times \left(\alpha  M-3 \sqrt{\alpha  (\alpha +1) M^2+r^2}\right)\Bigg\}^{-1} 
\end{eqnarray} 

\begin{figure}[ht!]
  \centering
   \includegraphics[width=0.98\linewidth]{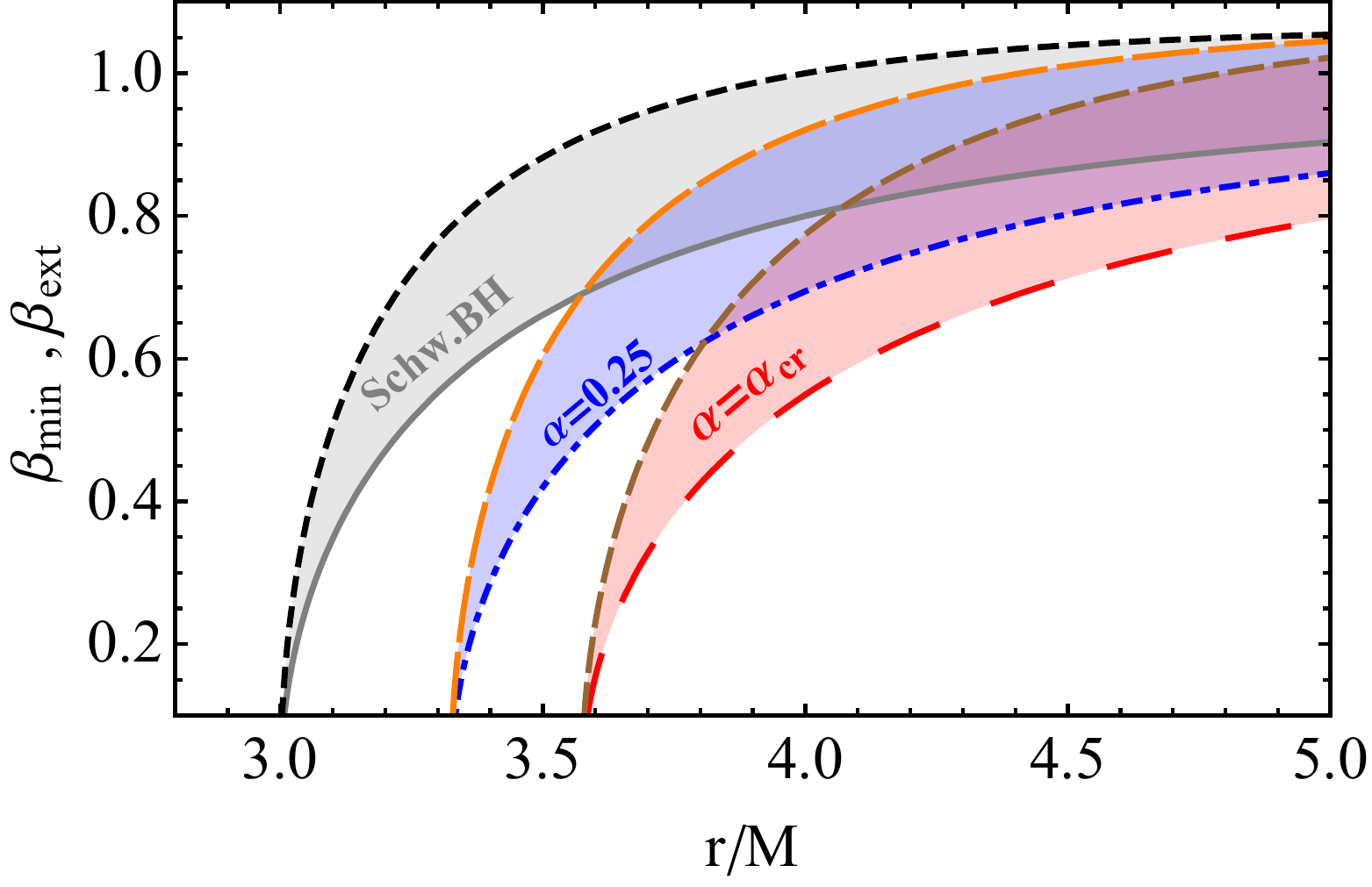}
    \caption{Radial profiles of extreme and the minimum magnetic coupling parameter of a magnetized particle at $l=0$ for different values of the MOG parameter $\alpha$. The shaded area implies the range of possible circular orbits for the fixed values of magnetic coupling parameter less than 1 ($\beta\leq \beta_{\rm exr}<1$) . \label{betaextremfig}}
\end{figure}

Figure \ref{betaextremfig} illustrates the radial dependence of the extreme value of the magnetic coupling parameter and the minimum value of the magnetic coupling parameter $\beta_{\rm min}~(l=0)$, for different values of the MOG parameter $\alpha$. Colored area implies the range of the parameter $\beta$ that allows the orbits of magnetized particles to be circular at $r\geq r_{min}$ or say, range of allowed circular orbits for a magnetized particle with the magnetic coupling parameter $\beta_{\rm extr}<\beta<\beta_{\rm min}(l=0)$ which can be stable. One can see from the figure that the inner circular orbits go outward when the MOG parameter is increased. To see effect of the MOG parameter on width of circular orbits for the magnetized particles with the parameter $\beta\geq \beta_{\rm extr}<1$ we have carried out numerical calculations of the system of equations 
\begin{equation}
    \beta = \beta_{\rm ext}(r_{\rm min},\alpha) , \quad \beta = \beta_{\rm min}(r_{\rm max},\alpha, l=0)\ ,
\end{equation}
and presented them table form for the different values of the magnetic coupling and MOG parameters. It is shown that both $\beta_{\rm min}(l=0)$ and $\beta_{\rm extr}$ are radially growing function. Here we calculate limits of values of the parameters   $\beta_{\rm min}(l=0)$ and $\beta_{\rm extr}$ at infinity
\begin{eqnarray}
\lim_{r \to \infty }\beta_{\rm extr}=\lim_{r \to \infty }\beta_{\rm min}(l=0)=1 \ .
\end{eqnarray}

It implies that a magnetized particle's stable circular orbits are placed at the infinity when the magnetic coupling parameter $\beta\geq 1$, in other words, the particle can not be in a circular stable orbit with $\beta \geq 1$ near the central black hole. In the case when a magnetized neutron star treated as a magnetized particle orbiting a SMBH, the limit $\beta <1$ can help to get a possible estimation of upper value for the external magnetic field around the black hole using observational parameters of the neutron star which can observe as recycled radio pulsars. For example, for a typical neutron star orbiting SMBH Sgr~A*,mass is 1.4 solar masses, a radius is 10 km and surface magnetic field is $10^{12}\rm G.$. One may assume that average magnetic field in close environment of the SMBH is about 100 G, calculations for the stable orbits show that the neutron star's dipolar magnetic field at the surface should be less than 1.623$ \times 10^{13}$ G in order to be in a circular orbit in the  close black hole environment. In the case of the magnetar SRG (PSR) J1745--2900 detected near the center of the Milky Way orbiting around Sgr~A* ~\cite{Mori2013ApJ} at the distance 0.1 pc from SrgA*, with magnetic dipole moment $\mu \approx 1.6\times 10^{32} \rm \ G\cdot cm^3$ and mass $m \approx 1.42 M_{\odot}$, the magnetic coupling parameter is about 7.16. It implies that the magnetar orbits can not be stable in the close environment of a SMBH.

\begin{table}[h!] \begin{center}\caption{\label{tab} Numerical values for the size of the area where circular orbits are allowed $\Delta r=r_{\rm max}-r_{\rm min}$ for different values of the magnetic interaction parameter $\beta$ and the parameter $\alpha$. Unit of $\Delta$ is given in $M$. } \begin{tabular}{|c| c| c| c| c| c| c| c| }\hline
$\alpha / \beta $ & $0$& $0.1$ & $0.3$ & $0.5$ & $0.7$ & $0.9$ & $1$ \\[1.ex]\hline %
$0 $ & 0 & 0.0019 & 0.0408 & 0.1573 & 0.3711 & 8.4872& $-$ \\[1.5ex]\hline
$0.1 $ & 6.2842 & 0.0038 & 0.0433 & 0.1434 & 0.3962 & 8.9282 & $-$ \\[1.5ex]\hline
$0.3 $ & 6.7514 & 0.0049 & 0.0482 & 0.1873 & 0.4455 & 9.7036 & $-$ \\[1.5ex]\hline
$ \alpha_{\rm cr} $ & 7.1481 & 0.0059& 0.0588 & 0.1969 & 0.5529 & 12.1403 & $-$  \\[1.5ex] \hline
\end{tabular} \end{center}
 \end{table}

Table~\ref{tab} demonstrates the range of circular orbits of the magnetized particles around the regular MOG black hole for different values of the magnetic interaction and parameter $\alpha$ of MOG.  One can see from the table~\ref{tab} that  with increase of the MOG and magnetic coupling parameters the area where stable circular orbits of magnetized particles are allowed becomes wider. At $\beta \geq 1$ the orbits can not be stable due to dominant effects of repulsive feature of magnetic interaction between magnetic dipole of the magnetized particle and the external magnetic field.  

\section{Conclusion \label{conclusion}}

In the present work we have studied the properties of the spacetime and event horizon of regular black hole in MOG gravity. It has been shown that the increase of the MOG parameter causes the decrease of scalar invariants of the spacetime of the black hole such as Ricci scalar, square of Riemann tensor and Kretchmann scalar. 

We have also explored the motion of neutral particles and obtained that positive (negative) values of the MOG parameter $\alpha$ causes the increase (decrease) of ISCO radius. We have shown similarity effects of MOG parameter on ISCO radius with effect of Kerr black hole and how the MOG parameter can mimic the spin of Kerr black hole providing the same values for ISCO radius of test particles. Energy efficiency of accretion disk by the model of Novikov and Thorne have been also studied and shown that the efficiency exceeds than 15 \%. 

The effect of MOG parameter on magnetic field in the vicinity of the regular BH has been also investigated. Study of charged particles motion has shown that ISCO radius of the charged particles around the black hole increases with the increase of MOG field due to behaviour of the field as additional gravity while the external magnetic field causes decrease of the ISCO radius. Moreover, we have also explored dynamics of magnetized particles around the regular MOG black hole in magnetic field in comoving frame of reference. The results indicate that the inner radius and the width of accretion disk consisting magnetized particles grows both with the increase of MOG parameter $\alpha$ and the increase of magnetic coupling parameter $\beta$. However, there are limitations  for $\beta$ outside of which there can be no stable circular orbits at a given radius and those boundaries decrease with increasing MOG parameter while the range of allowed values for magnetic coupling parameter $\beta$ is widened.

\section*{Acknowledgement}

This research is supported by Grants No. VA-FA-F-2-008, No.MRB-AN-2019-29 of the Uzbekistan Ministry for Innovative Development. AA acknowledges the support from Chinese academy of sciences through PIFI foundation.

\bibliographystyle{apsrev4-1}
\bibliography{gravreferences}
\end{document}